\documentclass[aps,pre,reprint,showpacs,showkeys,amsmath,amssymb]{revtex4-1}
\usepackage{graphicx}
\newcommand{\SO}[1]{\text{SO}(#1)}
\newcommand{\St}{\text{St}}
\newcommand{\USt}{\text{USt}}
\newtheorem{theorem}{Theorem}[section]
\newtheorem{definition}{Definition}
\allowdisplaybreaks

\begin{document}

\title{Deriving the Rosenfeld Functional from the Virial Expansion}
\author{Stephan Korden}
\email[]{stephan.korden@rwth-aachen.de}
\affiliation{Institute of Technical Thermodynamics, RWTH Aachen
University, Schinkelstra\ss e 8, 52062 Aachen, Germany}
\date{\today}

\begin{abstract}
In this article we replace the semi-heuristic derivation of the Rosenfeld
functional of hard convex particles with the systematic calculation of Mayer
clusters. It is shown that each cluster integral further decomposes into
diagrams of intersection patterns that we classify by their loop number. This
extends the virial expansion of the free-energy by an expansion in the loop
order, with the Rosenfeld functional as its leading contribution. Rosenfeld's
weight functions then follow from the derivation of the intersection
probability by generalizing the equation of Blaschke, Santalo, and Chern. It is
found that the 0-loop order can be derived exactly and reproduces the Rosenfeld
functional. We further discuss the influence of particle dimensions, topologies,
and geometries on the mathematical structure of the calculation.
\end{abstract}
\maketitle
\section{Introduction}\label{sec:introduction}
Hard particle systems serve as reference fluids for soft, granular, and cellular
matter. They interpolate the phase diagrams of molecular particles in the
limits of low and high particle densities, where the influence of the smooth and
attractive interactions is of secondary order. Phase transitions, as nematic
and smectic, can be understood as entropic effects of the excluded volume
\cite{hard-sphere-rev}. This distinguishes hard particle systems as the ideal
starting point for perturbation theory. However, it also requires an analytic
representation, or at least detailed knowledge, of the free-energy functional.
This requirement limits the usefulness of computer simulations, as the
minimization procedure needs the functional form of the free-energy and not its
function. To obtain a theoretical understanding of the liquid, crystalline,
amorphous, and glassy states \cite{stillinger} we therefore need better
analytical tools than are currently available.

During the last decades, several interesting approximations have been developed
to derive analytical expressions of the free-energy or the pair-correlation
function \cite{mcdonald}. However, most of them are restricted to hard spheres,
such as the well known solution of Thiele and Wertheim \cite{thiele,
wertheim-py1, wertheim-py2} of the Ornstein-Zernicke equation in the
Percus-Yevick approximation. A different approach was suggested by Reiss,
Frisch, and Lebowitz \cite{scaled-particle-1}, who used the result of Isihara
and Kihara \cite{isihara-orig, kihara-1, kihara-2} for the second virial
coefficient for convex particles. Their scaled particle theory motivated
Rosenfeld \cite{rosenfeld-structure, rosenfeld-freezing, rosenfeld-closure,
rosenfeld-mixture, rosenfeld1} to develop the fundamental measure theory, which
is based on the local decoupling of the second virial integral, on the
invariance of the free-energy functional under coordinate rescaling and on its
solution of the scaled particle differential equation. The Rosenfeld functional
is therefore a semi-heuristic result, valid under the same assumptions as the
scaled particle theory. Nevertheless, its advantage is the explicit dependence
on the particle geometry through the weight functions and its local
representation of the free-energy functional as the convolute of weight
densities.

Later it has been shown by Rosenfeld and Tarazona \cite{tarazona-rosenfeld,
tarazona, crossover-rosenfeld-1, crossover-ros-1, crossover-rosenfeld-2} that
the functional leads to an inconsistency when the volume, filled with spheres,
is restricted to a single layer, a tube or a one-particle cavern. This led to a
geometrically motivated correction of the original form and resulted in a
highly accurate functional for the fluid phase of hard spheres up to the
freezing point \cite{tarazona, comparision-ros}. A different strategy made use
of simulation data to go beyond the Percus-Yevick approximation
\cite{white-bear-1, white-bear-2}. The simple structure of the functional led to
further applications for cylinders, discs, needles, and their mixtures
\cite{schmidt-dft, schmid-c, schmidt-mixtures, schmidt-colloidal,
schmidt-one-dim, schmidt-palelet, goos-mecke} and alternative representations
of the weight functions \cite{rosinberg-kierlik-1, rosinberg-kierlik-2}. For a
recent review see also \cite{roth_rev}. However, despite its success, it is not
clear, how to extend Rosenfeld's approach further and how to go beyond the
semi-heuristic construction of the functional.

In this article we will begin an investigation to clarify the underlying
mathematical and physical assumptions of the fundamental measure theory. In a
first step it will be shown that the Rosenfeld functional is only the leading
order of an infinite expansion of the free-energy in intersection diagrams,
which will be classified by their number of loops and intersection centers. The
0-loop order corresponds to sets of particles that intersect in at least one
common point and can freely rotate around this center. This intersection pattern
corresponds to an infinite subset of Mayer clusters and will be derived in this
work by generalizing the equation of Blaschke, Santalo, and Chern
\cite{blaschke, santalo-book, chern-1, chern-2, chern-3}. It will be shown that
the infinite number of terms of the 0-loop contribution requires only the
calculation of three Euler forms. The relation between the intersection
probabilities and their corresponding subsets of Mayer clusters then allows the
calculation of the free-energy functional via the virial expansion. However,
instead of the virial series in the single-particle density, we have to
interpret the expansion in Rosenfeld's weight densities. This reformulation of
the virial expansion not only reproduces the Rosenfeld functional as the
0-loop order but applies also to all further loop orders. 

The article is divided into two sections. The scope of part
\ref{sec:integral_geometry} is more general. Here, we introduce the concept of
the loop expansion, \ref{subsec:stacks}, give some background information on
differential and integral geometry, \ref{subsec:topology}, and lastly derive the
weight functions from integral geometry in section \ref{subsec:branching}. Part
\ref{sec:rosenfeld} considers the 0-loop contribution of the free-energy.
\ref{subsec:ros-tara} recapitulates Rosenfeld's ideas leading to the
semi-heuristic formulation of the fundamental measure theory and Tarazona's
corrections.  This approach is compared in \ref{subsec:ros_functional} to our
new ansatz via the virial expansion, where we derive the 0-loop contribution of
the free-energy and prove its equivalence to Rosenfeld's functional. We end the
article in \ref{sec:conclusion} with a discussion of the convergence of the loop
expansion.
\section{The Intersection Probability of Particle Stacks}
\label{sec:integral_geometry}
\subsection{Intersecting Particle Stacks}\label{subsec:stacks}
So far, little attention has been paid to the approximation scheme leading to
Rosenfeld's free-energy functional. Instead, fundamental measure theory is based
on a tower of three postulates that fix the functional's overall structure:
1. the free-energy functional density is assumed to be a polynomial in weight
densities, 2. uniquely determined by its the homogeneous scaling dimension and
3. its solution of the empirical scaled particle differential equation.
Actually, there is no physical argument that justifies these assumptions from
first principles, and the failure of only one postulate could cause the downfall
of the remaining parts. A first step in generalizing Rosenfeld's approach is
therefore to test these three postulates and, if necessary, to replace them.
This will be done in the following by comparing the third virial order of
Rosenfeld's functional to its exact integral. 

Rosenfeld's truly remarkable step in developing a weighted density functional
for spheres is the local splitting of Mayer's f-function into weight functions
and to recognize its relation to the Gauss-Bonnet equation and thus to the
Gaussian curvature $K$ \cite{rosenfeld-structure, rosenfeld2, rosenfeld-gauss2}.
For two particles $D_i, D_j$, intersecting in a domain $A = D_i\cap
D_j$ of coordinate vector $\vec{r}_A \in D_i\cap D_j$, the f-function decouples
into the convolute
\begin{align}\nonumber
&f_{ij}(\vec{r}_{ij}) 
= -\frac{1}{4\pi}\int_{\partial (D_i\cap D_j)} K(A)\, dS_A \\
&\qquad\quad \, =-\frac{1}{4\pi}\int_{D_i\cap D_j}\quad 
K(A)\,\delta(\vec{n}\vec{r}_A) \, d^3r_A\label{f-function_1}\\
&=\int_{D_i\cap D_j}C^{A_1A_2}
\omega^i_{A_1}(\vec{r}_A)\omega^j_{A_2}
(\vec{r}_A-\vec{r}_{ij})\, d^3r_A\nonumber\\
&=\int_{D_i\cap D_j}C^{A_1A_2}
\omega^i_{A_1}(\vec{r}_A-\vec{r}_i)\omega^j_{A_2}
(\vec{r}_A-\vec{r}_j)\, d^3r_A\label{f-function}
\end{align}
depending on the particle positions $\vec{r}_i\in D_i$, $\vec{r}_j\in
D_j$ and the distance vector $\vec{r}_{ij}=\vec{r}_j-\vec{r}_i$. In
(\ref{f-function_1}) the integration over the surface $S_A$ has been
transformed into a volume integral at normal vector $\vec{n}$ by equation
(\ref{d-1}) and finally arranged in the symmetric form (\ref{f-function}),
assuming that the embedding space of the particles is of infinite volume.

The derivation of the local decoupling of the f-function (\ref{f-function}) and
its relationship to the Gauss-Bonnet equation will be explained in the following
sections and deduced from the Blaschke, Santalo, Chern equation of integral
geometry \cite{blaschke, santalo-book, chern-1, chern-2}. It provides an exact
identity for the intersection probability of convex particles and determines
the pre-factor $1/(4\pi)$ uniquely. Based on this result, any Mayer diagram can
be transferred into weight functions.

The third virial cluster in this representation is now an integral over three
intersection centers $A,B,C$, particle positions $\vec{r}_i$, $\vec{r}_j$,
$\vec{r}_k$, and their corresponding rotations $\vec{\Omega}$. Let us introduce
the notation 
\begin{equation}\label{def-measure}
\begin{split}
\Gamma(D):= \{\, \gamma = & \;(\vec{r} \,,\, \vec{\Omega} )\; |\; \vec{r}\in
D\,,\, \vec{\Omega}\in \SO{3}\,\}\\[0.25em]
d\gamma_i &:= d^3r_i\, d^3\Omega_i
\end{split}
\end{equation}
for the differential volume element. The exact third virial integral has thus
the form:
\begin{equation}\label{3-exact}
\begin{split}
\beta_2^{(1)}&=\frac{1}{2V} \int \;
C^{A_1A_2}\omega^i_{A_1}(\vec{r}_A-\vec{r}_i)
\omega^j_{A_2}(\vec{r}_A-\vec{r}_j)\\[0.4em]
&\qquad\quad\times
C^{B_1B_2}\omega^j_{B_1}(\vec{r}_B-\vec{r}_j)
\omega^k_{B_2}(\vec{r}_B-\vec{r}_k)\\[0.4em]
&\qquad\quad \times \;C^{C_1C_2}\omega^k_{C_1}(\vec{r}_C-\vec{r}_k)
\omega^i_{C_2}(\vec{r}_C-\vec{r}_i)\\[0.4em]
&\qquad\quad  \times
\delta(\vec{r}_{AB}+\vec{r}_{BC}+\vec{r}_{CA})\;\,d^3r_A
d^3r_B d^3r_C\\[0.4em]
&\qquad\quad  \times
\rho_i(\vec{r}_i)\rho_j(\vec{r}_j)\rho_k(\vec{r}_k)\;\,d\gamma_i d\gamma_j
d\gamma_k
\end{split}
\end{equation}
restricted by the loop constraint
\begin{equation}\label{loop}
\vec{r}_{AB}+\vec{r}_{BC}+\vec{r}_{CA}=0
\end{equation}
of their distance vectors $\vec{r}_{AB}=\vec{r}_B-\vec{r}_A$. Collecting terms
according to their particle number and introducing Wertheim's 2-point density 
\cite{wertheim-1, wertheim-2,wertheim-3, wertheim-4}
\begin{equation}\label{2-point}
\begin{split}
\bigl<\omega^i_{A} & \omega^i_{B}\rho_i\bigr>(\vec{r}_{AB}) \\
&= \sum_i \int_{\Gamma(D_i)}
\omega^i_A(\vec{r}_A-\vec{r}_i)
\omega^i_B(\vec{r}_B-\vec{r}_i)\rho_i(\vec{r}_i)\,d\gamma_i
\end{split}
\end{equation}
for $\vec{r}_A, \vec{r}_B \in D_i$, equation (\ref{3-exact}) can be written in
the more symmetric form:
\begin{equation}\label{3-exact-2}
\begin{split}
\beta_2^{(1)}
&=\frac{1}{2V} C^{A_1A_2} C^{B_1B_2} C^{C_1C_2}\\
&\times 
\int \bigl<\omega^i_{A_1}\omega^i_{C_2}\rho_i\bigr>(\vec{r}_{CA})
\bigl<\omega^j_{B_1}\omega^j_{A_2}\rho_j\bigr>(\vec{r}_{AB})\\
&\quad \times
\bigl<\omega^k_{C_1}\omega^k_{B_2}\rho_k\bigr>(\vec{r}_{BC})\\[0.4em]
&\quad \times \;
\delta(\vec{r}_{AB}+\vec{r}_{BC}+\vec{r}_{CA})\;\,d^3r_{AB}
d^3r_{BC} d^3r_{CA}\;.
\end{split}
\end{equation}
This is to be compared to the corresponding third virial integral obtained from
Rosenfeld's functional:
\begin{equation}\label{3-approx}
\begin{split}
&\beta_2^{(0)} = \frac{1}{V}\int 
\Bigl[ \frac{1}{2}\omega^i_\chi(\vec{r}_A-\vec{r}_i)
\omega^j_v(\vec{r}_A-\vec{r}_j) \omega^k_v(\vec{r}_A-\vec{r}_k)\Bigr.\\
&\;+\;\;\;C^{\alpha_1\alpha_2}\;\;\omega^i_{\alpha_1}(\vec{r}_A-\vec{r}_i)
\omega^j_{\alpha_2}(\vec{r}_A-\vec{r }_j)
\omega^k_v(\vec{r}_A-\vec{r}_k)\\[0.4em]
&\; +C^{\alpha_1\alpha_2\alpha_3}
\omega^i_{\alpha_1}(\vec{r}_A-\vec{r}_i)
\omega^j_{\alpha_2}(\vec{r}_A-\vec{r}_j)
\Bigl.\omega^k_{\alpha_3}(\vec{r}_A-\vec{r}_k)
\Bigr]\\[0.4em]
&\qquad\quad \times \rho_i(\vec{r}_i)\rho_j(\vec{r}_j)
\rho_k(\vec{r}_k)\;\,d\gamma_i d\gamma_j d\gamma_k\;\,d^3r_A\;,
\end{split}
\end{equation}
which has a much simpler form, integrated over only one intersection center $A
\subset D_i\cap D_j \cap D_k$ and depends on three weight functions only,
instead of six in the exact expression (\ref{3-exact}). 

\begin{figure}
\includegraphics[width=3cm,angle=-90]{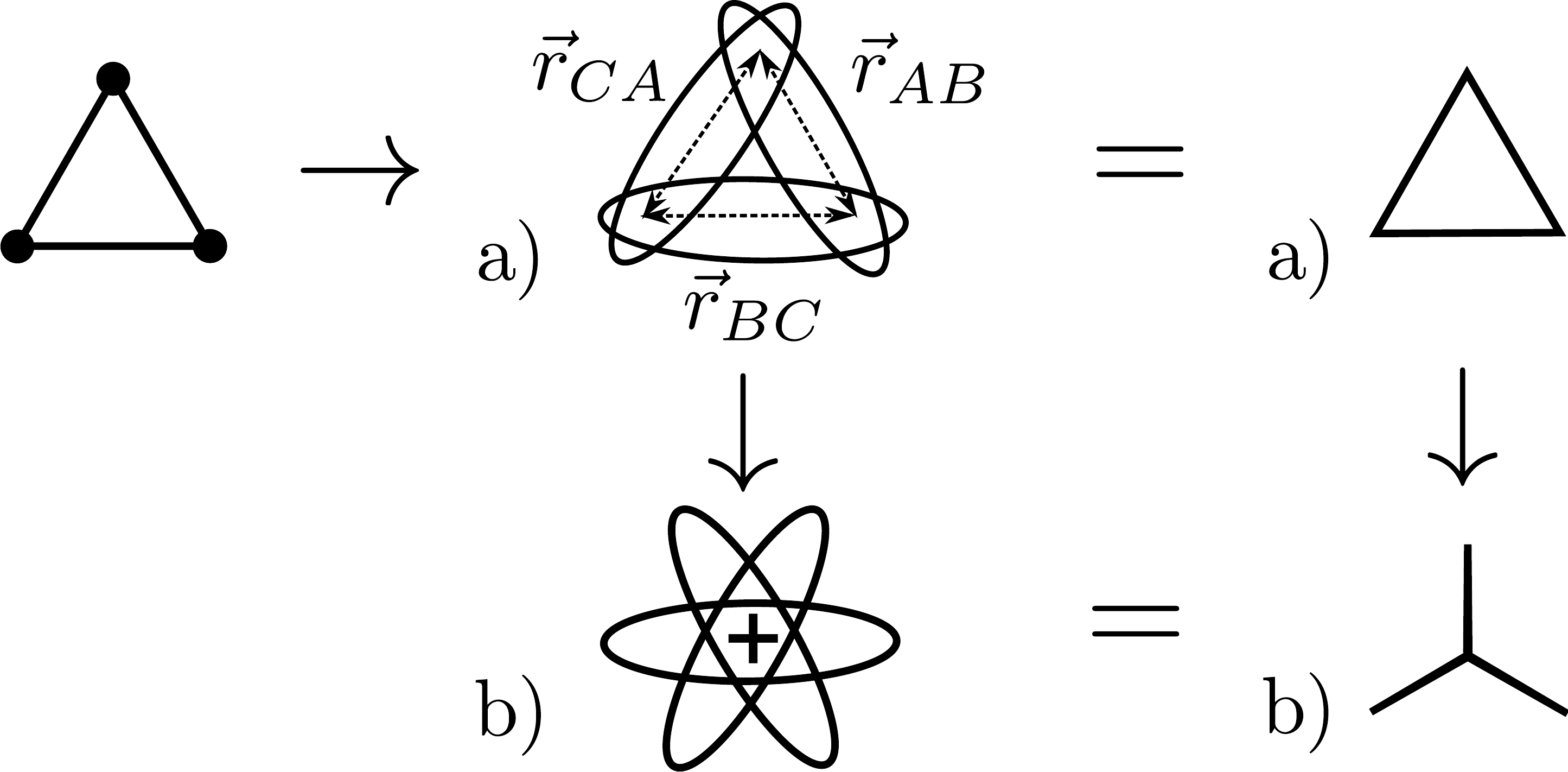}
\caption{The third virial Mayer diagram in the particle (left) and intersection
representation (right): a) pairwise intersecting particles corresponding to the
exact cluster integral and b) its approximation as the stack of third order.}
\label{fig:3-virial}
\end{figure}

The principal difference between these two integrals is illustrated in
FIG.~\ref{fig:3-virial}.
FIG.~\ref{fig:3-virial}a) displays the generic intersection pattern of the third
virial integral with pairwise overlapping domains (\ref{3-exact}), whereas the
corresponding figure FIG.~\ref{fig:3-virial}b) shows the case of
(\ref{3-approx}) with only one such center. Rosenfeld's diagram is a degenerate
third virial coefficient, obtained in the limit $\vec{r}_A =\vec{r}_B
=\vec{r}_C$, where the triangle of FIG.~\ref{fig:3-virial}a) shrinks to the tree
diagram of FIG.~\ref{fig:3-virial}b). The difference between the exact and the
approximated third virial integral is therefore the way in which the particles
intersect each other. 

Instead of the graphical representation of intersecting particle domains, it is
sufficient to symbolize the intersection patterns in ``intersection diagrams'',
where the particles correspond to lines and intersection centers to the
position where the lines join. The corresponding diagrams of the third virial
are shown on the right of FIG.~\ref{fig:3-virial}. 

\begin{figure}
\includegraphics[width=1.2cm,angle=-90]{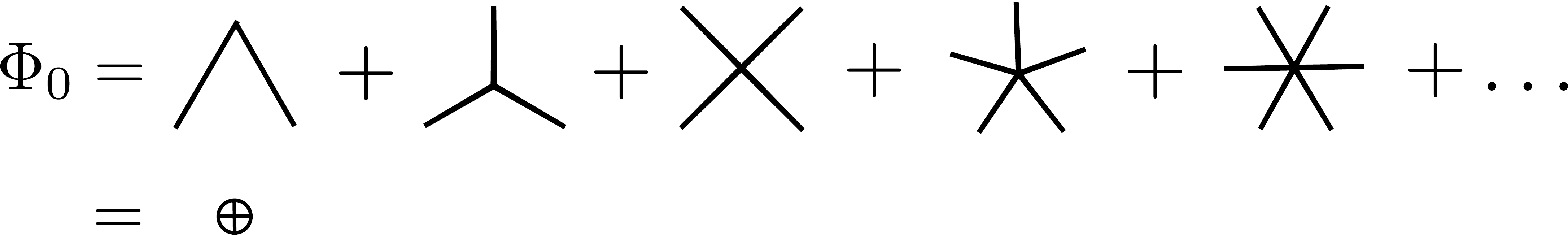}
\caption{The Rosenfeld functional is the 0-loop approximation of the
free-energy. Each intersection diagram corresponds to a completely connected
Mayer cluster, contracted into a stack. The sum over all such diagrams is
symbolized by a crossed circle.}
\label{fig:zero-loop}
\end{figure}

Rosenfeld's functional contains an infinite number of further virial
contributions. These are obtained by Taylor expanding the singular parts in
powers of the weight function $\omega_v$ and have the generic form:
\begin{equation}
C^{\alpha_1\alpha_2\alpha_3}[\omega_{\alpha_1}\omega_{\alpha_2}\omega_{\alpha_3}
(\omega_v)^{n-3}](\vec{r}_A; \vec{r}_1,\ldots, \vec{r}_n)
\end{equation}
corresponding to Mayer diagrams, whose intersection domains have been contracted
into one single domain, as shown in FIG.~\ref{fig:zero-loop}. 
However, only completely connected Mayer clusters interact in such a way that
each particle interacts with each other. Rosenfeld's functional is therefore
the sum over an infinite number of completely connected diagrams that are
further contracted into one intersection point.

The arguments, obtained so far, can be summarized in the following way: The
exact free-energy functional is not representable by Rosenfeld's weight
densities alone. Instead, the third virial integral (\ref{3-exact-2}) is a
function of Wertheim's 2-point densities (\ref{2-point}), and it is natural to
assume that this result has to be generalized to arbitrary k-point densities.
Next, as the Mayer function (\ref{f-function}) is itself invariant under
coordinate scaling, it is not possible to restrict the functional form by its
scaling dimension. From this follows that the three postulates of FMT, including
the empirical scaled particle differential equation, have no deeper physical
basis. On the other hand, we have also seen that Rosenfeld's functional
approximates and re-summes a certain class of Mayer diagrams contracted to one
intersection point, as shown in FIG.~\ref{fig:3-virial}. This offers an
alternative approach to derive the functional and, most importantly, it also
opens a path to derive higher order corrections. 

The central object of FMT is the sum of contracted intersection diagrams, shown
in FIG.~\ref{fig:zero-loop}. Because of its importance, let us introduce the
name ``stack'' for individual parts and ``universal stack'' for its sum, defined
by:
\begin{definition}\label{stack}
A stack of order $k=\text{ord}(\St_k)$ is a set of $i=1,\ldots, k$ domains
$D_i$, intersecting in at least one common point and free to translate and
rotate around this center: 
\begin{equation}
\St_k = \bigcap_{i=1}^k D_i\;.
\end{equation}
The universal stack is the formal sum over all stacks intersecting at the same
point
\begin{equation}
\USt = \bigoplus_{i=2}^\infty \St_k\;.
\end{equation}
\end{definition}
In the following sections we will prove that the intersection probability of the
universal stack $\Phi_0$ reproduces Rosenfeld's functional $\Phi_{\text{R}}$
\begin{equation}\label{aim}
\Phi_R = \Phi_0\;.
\end{equation}

However, this is only the first hint to a more general structure: When
completely contracted intersection diagrams correspond to a free-energy
functional at low packing fraction, it is natural to assume that diagrams, not
completely contracted, provide higher order corrections. 

\begin{figure}
\includegraphics[width=3.5cm,angle=-90]{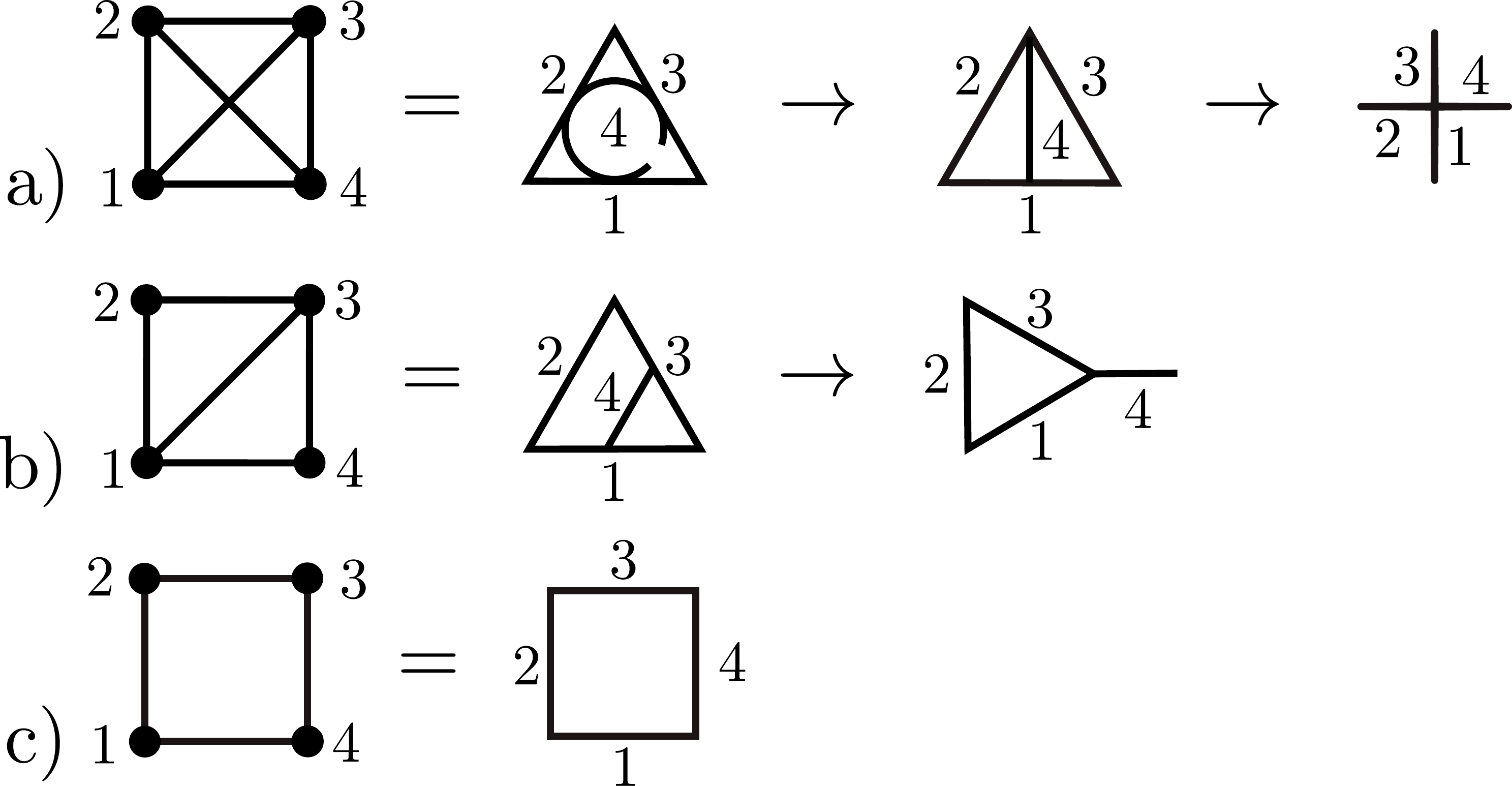}
\caption{Mayer clusters of the fourth virial order, translated into intersection
diagrams and ordered by the tuple $(g,n)$: a) $(3,6)$, $(2,4)$, $(0,1)$, b)
$(2,5)$, $(1,3)$, and c) $(1,4)$.}
\label{fig:four-virial}
\end{figure}

Rosenfeld's functional is exact for the second virial order. The third
virial integral, however, is only an approximation, as shown in
FIG.~\ref{fig:3-virial}. Adding the exact third virial diagram will therefore
result in an improved functional, corresponding to three additional intersection
centers and the loop constraint (\ref{loop}). In principle, it is possible to
add arbitrary intersection diagrams to the functional, systematically derived
from the Mayer clusters. As an example consider FIG.~\ref{fig:four-virial}
where the 4-particle Mayer diagrams are shown together with their corresponding
set of intersection diagrams and contractions, ordered by their number of loops
$g$ and intersection centers $n$. 
This classification by the tuple $(g,n)$ comes natural as the calculational
complexity increases with both. However, they have also a direct physical
interpretation.

\begin{figure}
\includegraphics[width=1.3cm,angle=-90]{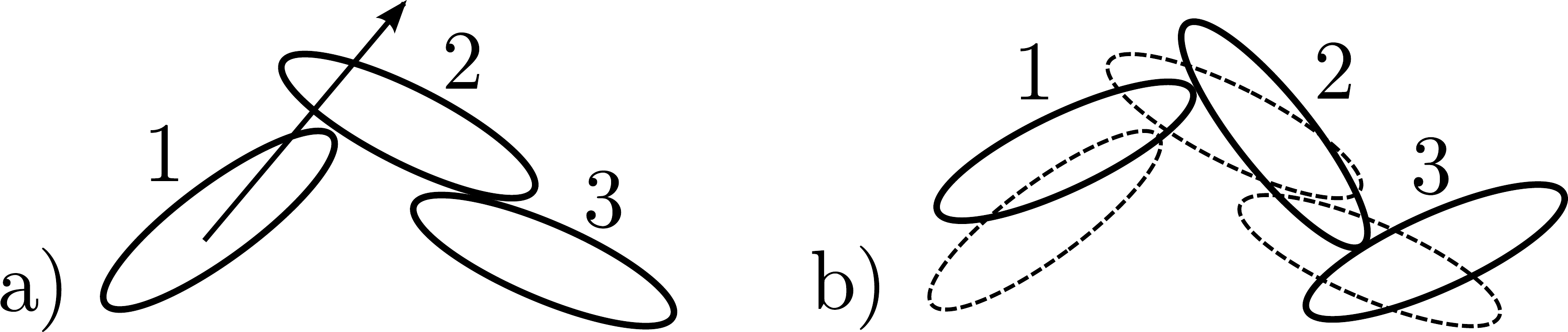}
\caption{The displacement of one particle causes a shift in the position of
all neighbors that are in direct and indirect contact, resulting in long range
correlations between particles.}
\label{fig:blocking}
\end{figure}

The loop order $g$ counts the number of constraints, restricting the coordinates
of intersection domains:
\begin{equation}\label{loop-general}
\begin{split}
\left. 
\begin{aligned}
\vec{r}_{A_1A_2}+ &\ldots + \vec{r}_{A_{k-1}A_k} = 0\\
\vec{r}_{B_1B_2}+ &\ldots + \vec{r}_{B_{l-1}B_l}\, = 0\\
&\ldots
\end{aligned}
\right\}
\quad \text{$g$ loop constrains}
\end{split}
\end{equation}
for $g$ loops with $k, l, \ldots$ intersection centers. In this way,
correlations are generated between particles that otherwise do not
interact directly via a potential function. This distinguishes the zero loop
order $g=0$, where no such constraints exists, providing a plausible
argument why Rosenfeld's free-energy functional describes only the fluid
regime below the first phase transition and predicts a maximum packing fraction
at $n_v=1$, independent of the particle geometry. The solid phase region, on the
other hand, requires long range correlations between particles, such that
shifting one particle leads to the displacement of others, as shown in
FIG.~\ref{fig:blocking}.

\begin{figure}
\includegraphics[width=1.9cm,angle=-90]{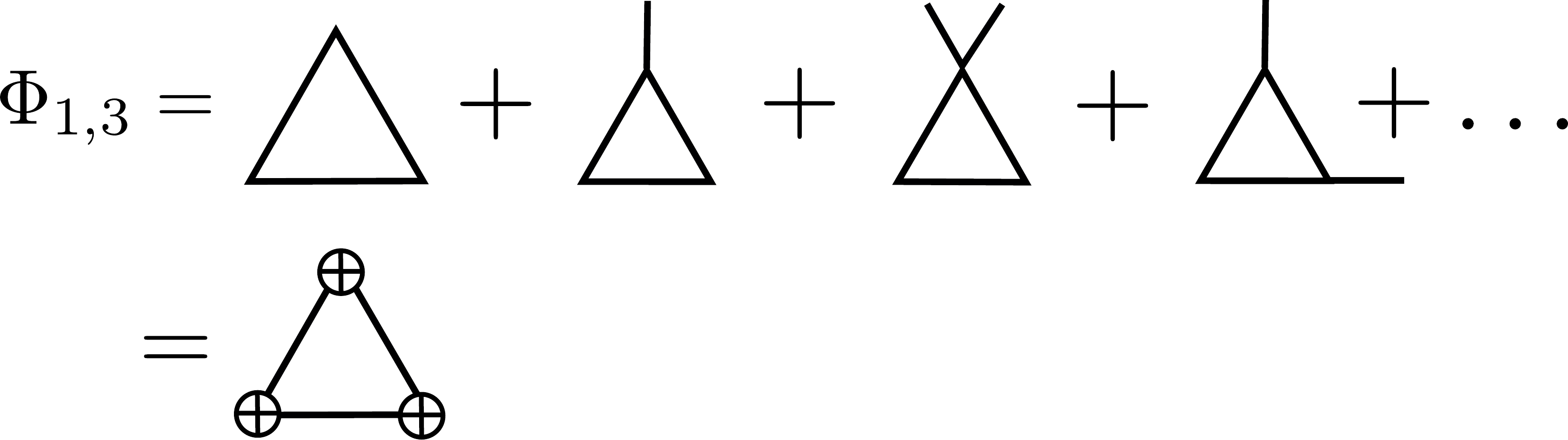}
\caption{The re-summed and regularized third virial integral: Adding additional
particles to an existing intersection point does not increase the calculational
effort.}
\label{fig:third-virial-reg}
\end{figure}

These considerations make the number of loops $g$ and intersections $n$
convenient indices to group the diagrams and to define the ``loop expansion'' of
the free-energy excess functional:
\begin{equation}\label{top-exp}
\Phi^{\text{ex}} = \sum_{g=0}^\infty (\sum_{n=1}^\infty \Phi_{g,n}) = 
\sum_{g=0}^\infty \Phi_g\;,
\end{equation}
where each element $\Phi_{g,n}$ corresponds itself to an infinite number of
intersection diagrams. Examples are shown for $\Phi_{0,1} = \Phi_0$ in
FIG.~\ref{fig:zero-loop} and for $\Phi_{1,3}$ in
FIG.~\ref{fig:third-virial-reg}. 
It is worth pointing out that some of the contracted 4-particle diagrams
of FIG.~\ref{fig:four-virial} turn up as corrections of the second and third
virial order. Actually, it will be shown in the following sections that the
calculational effort does not increase when additional particles are added to
an existing intersection point. Any individual intersection diagram can
therefore be replaced by a ``resummed diagram'', with each intersection point
replaced by the universal stack. Resummation is therefore a central aspect of
FMT, as it generates the pole structure in the free volume $1-n_v$, which is so
characteristic for Rosenfeld's functional. 

\begin{figure}
\includegraphics[width=0.9cm,angle=-90]{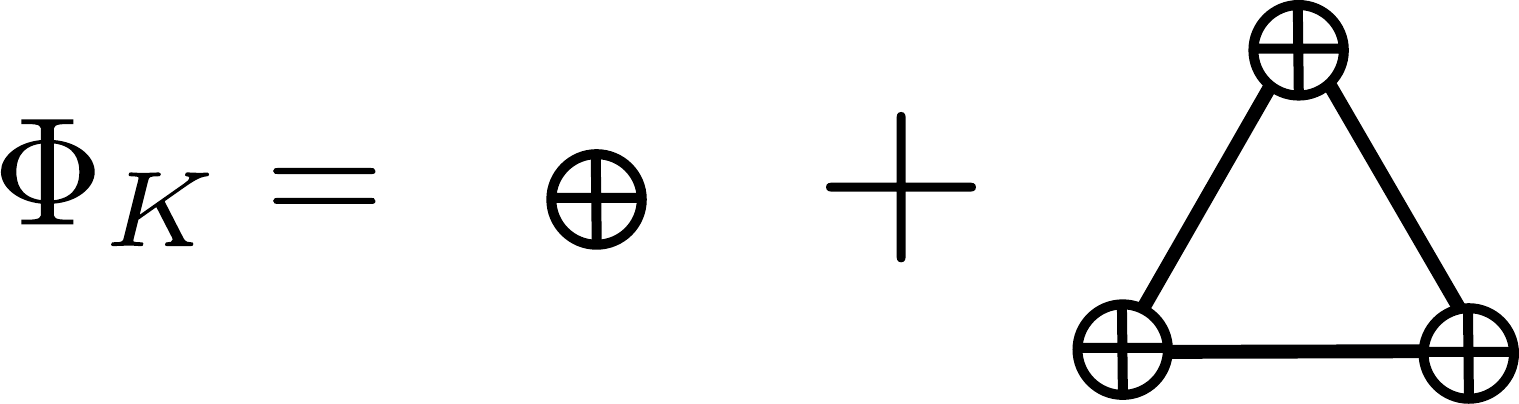}
\caption{Going beyond Rosenfeld's functional: A first approximation for the
1-loop order contains the re-summed second virial integral and the regularized
third virial integral.}
\label{fig:beyond-ros}
\end{figure}

A natural extension of the current functional is the combination $\Phi_K =
\Phi_{0,1}+\Phi_{1,3}$, shown in FIG.~\ref{fig:beyond-ros}. 
However, as parts of $\Phi_{0,1}$ are already included in $\Phi_{1.3}$, it is
necessary to ``regularize'' the loop diagram by excluding the case
$|\vec{r}_{AB}| = |\vec{r}_{BC}| = |\vec{r}_{CA}|=0$, shown in 
FIG.~\ref{fig:3-virial}, where the distances between the intersection
coordinates vanish. All loop diagrams are understood in this way, excluding the
case of collapsing loops and thus ensuring that regularized diagrams are
uniquely defined.

\begin{figure}
\includegraphics[width=0.7cm,angle=-90]{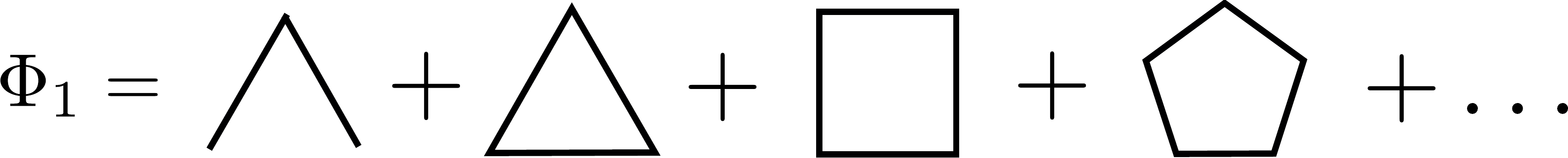}
\caption{The unregularized free-energy at 1-loop order is the sum over the
second virial and all further ring diagrams.}
\label{fig:1-loop-gen}
\end{figure}

Apart from the resummation of intersection points, it is also possible to sum
up diagrams of identical loop order. One example is displayed in
FIG.~\ref{fig:1-loop-gen}. The analytical structure of the generating function
$\Phi_1$ can be derived from the virial expansion,
as the ring diagrams are formally identical to Mayer clusters. With the symmetry
factor $(k-1)!/2$ for a ring of $k$ particles and the simplifying notation
$C^{AB}\omega_A\omega_B$ for the f-function (\ref{f-function}), the 1-loop
free-energy yields the formal expression:
\begin{equation}
\begin{split}
\Phi_1 & =  \Bigl<\sum_{k=1, k\neq 2}^\infty \frac{1}{k!}\frac{1}{2}(k-1)!
\bigl( C^{AB}\omega_A\omega_B\rho \bigr)^k \Bigr>\\
& = - \frac{1}{2}\Bigl<\ln{\bigr(1- C^{AB}\omega_A\omega_B\rho \bigr)}\Bigr>
+ \ldots \;,
\end{split}
\end{equation}
where the angular brackets indicate the integration over the coordinates. The
1-loop free-energy contribution is therefore of a completely different structure
than Rosenfeld's functional, signaling a logarithmic divergence, depending on
the particles' geometry.

Having identified the approximation scheme behind Rosenfeld's functional, we
will now begin with the development of the mathematical framework necessary
to derive the intersection probability of the universal stack. In this way,
the hypothesis (\ref{aim}) will be proven by direct calculation, which is the
basis for the resummation of intersection points and all further constructions
that will be considered in following papers. 
\subsection{Some Relevant Information on Differential Geometry}
\label{subsec:topology}
\subsubsection{Intrinsic Geometry}
The derivation of the intersection probabilities requires the introduction of
some mathematical conventions \cite{kobayashi, guggenheimer,
complex-manifolds-potential} and the discussion of physical constraints. 

\begin{figure}
\centering
\includegraphics[width=1.7cm,angle=-90]{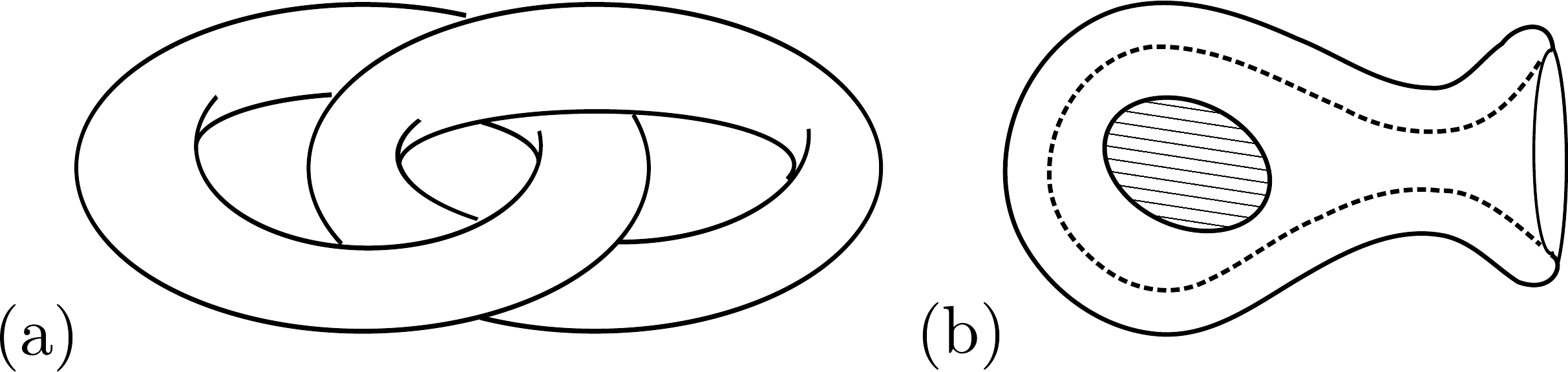}
\caption{Physical constraints restrict the regions accessible for other
particles. Forbidden configurations are: a) linked tori and b) particles
inside a cavern (bottle) with pore opening smaller than the particles'
diameter.}
\label{fig:forbidden}
\end{figure}

Let $D$ denote an Euclidean, Riemannian manifold of $3$ dimensions, sufficiently
differentiable to allow for the calculation of the Euler form. Manifolds of this
type include a variety of geometries as convex and concave particles, Klein's
bottle, tori, polyhedrons, cylinders, hollow spheres but also non-compact
structures. The mathematical requirements are therefore not very restrictive.
However, we also have to take into account the physical constraints. In the
formulation of Mayer's f-functions, the cluster integrals determine the
intersection probability between particles. However, the physical
particle domain is not only restricted by its surface. Instead one has to
determine the region that is inaccessible for other particles.
FIG.~\ref{fig:forbidden} shows two examples, where the corresponding
mathematical intersection probability is zero but not its physical one.
FIG.~\ref{fig:forbidden}a) shows two linked tori. In a fluid of single tori,
such a configuration has to be excluded as the particles cannot penetrate
each other. The same applies to the system of a particle inside a concave
domain, whose opening is smaller than the particle's smallest diameter, as
seen in FIG.~\ref{fig:forbidden}b). When the geometry of the first does not
allow to enter the inner region of the second, it has to be excluded, i.e.
counted as part of the domain of the latter particle. 

Although physically related, the mathematical nature of these two examples is
very different. The case of two tori is related to Euler's linking number
\cite{bott-tu, nash} and belongs to the topological class of homotopically
non-trivial intersections. Another example is the intersection of hollow spheres
as realized by fullerenes. Both cases are related to topological classes that
follow by successive variations of Euler forms \cite{alvarez-ginsparg}. And
although they are not required for the current article, they are interesting
enough to give a short account further below. The second case  
FIG.~\ref{fig:forbidden}b) is more difficult to solve. Here, we have to
introduce a fictitious membrane at the opening of the pore, whose surface vector
is always antiparallel to the surface vector of the docking particle. Such
configurations lead to the vanishing of certain contributions of the
intersection probability between particles, as we will show in the next section,
and might give new insight into the isotropic-nematic phase transition. Because
of these additional complications, we will exclude homotopically non-trivial
particles as well as concave geometries. The discussion simplifies further, when
boundaries are excluded, leaving us with 3-dimensional convex particles embedded
into the flat Euclidean space $\mathbb{R}^3$. 

The geometry of a physical particle depends on intrinsic and extrinsic
properties, i.e. the properties independent and dependent on the embedding.
It would be therefore sufficient to consider 2-dimensional surfaces and their
embedding into $\mathbb{R}^3$. However, at this point it is worthwhile to
discuss Cartan's formulation of differential geometry \cite{bruhat, kobayashi,
guggenheimer} for general dimension, as some of the results degenerate for low
dimensional spaces.

Let the particle $\Sigma$ be a $n$ dimensional, orientable, differential
Riemannian manifold without boundary. Suppose further that the manifold can be
covered by a set of open coordinate patches $\Sigma=\cup_\alpha U_\alpha$, 
each one isomorphic to $\mathbb{R}^n$ and labeled by a local, orthonormal
coordinate frame $(p,e_1^{(\alpha)}, \ldots, e_n^{(\alpha)})$ at the point
$p\in U_\alpha$. The local frames at overlapping regions $U_\alpha\cap U_\beta$
are related to each other by differentiable coordinate transformations
$g_{\alpha \beta}(p): U_\alpha \cap U_\beta \to \SO{n}$. The matrix valued
transition functions $g_{\alpha \beta}$ are invertible $g_{\alpha\beta}^{-1} =
g_{\beta \alpha}$ and fulfill the cyclic condition $g_{\alpha\beta}
g_{\beta\gamma} g_{\gamma\alpha} =1$ at triple intersections $U_\alpha\cap
U_\beta \cap U_\gamma$. These preliminaries define the tangential bundle
$T\Sigma$ with the local section $(p, e_1, \ldots, e_n) \in \Gamma (T\Sigma)$
and the cotangential bundle $T^*\Sigma$ as its dual space, related to $T\Sigma$
by the metric of
$\mathbb{R}^n$ 
\begin{equation}\label{metric}
e_i\cdot e_j = \eta_{ij}
\end{equation}
and its differential structure. The vielbein $\theta_i$ and connection forms
$\omega_{ij}$ are defined by
\begin{equation}\label{cartan-1}
dp = e_i\theta_i \;, \quad de_i = \omega_{ij}e_j \quad \text{for} \quad
p\in \Sigma
\end{equation}
and transform under the coordinate change $e_i' = g_{ij} e_j$ as
\begin{equation}\label{transform-1}
\theta_i' = g_{ij}\theta_j \;, \quad 
\omega_{ij}' = g_{ki}^{-1}\omega_{kl}g_{lj} + g_{ki}^{-1}dg_{kj}\;,
\end{equation}
where summation over paired indices is understood. The connection is therefore
not a tensor and can be locally replaced by a trivial gauge. 

The vanishing of the second exterior derivative of ({\ref{cartan-1}) defines the
torsion and the curvature form
\begin{equation}\label{cartan-2}
T_i = d\theta_i - \omega_{ij}\wedge \theta_j \;,\quad 
\Omega_{ij} = d\omega_{ij} - \omega_{ik}\wedge \omega_{kj}\;,
\end{equation}
which transform as a first and second rank tensor
\begin{equation}\label{transform-2}
T_i' = g_{ij}T_j \;, \quad 
\Omega'_{ij} = g_{ki}^{-1}\Omega_{kl}g_{lj}\;.
\end{equation}
The constraint $T_i=0$ of a Riemannian manifold is therefore independent of the
coordinate system and introduces a global relationship between the vielbein and
the connection forms. 

Torsion and curvature carry local information about the geometry of a manifold,
always restricted to single coordinate patches $U_\alpha$ and depending on the
chosen coordinate system. Globally defined forms, on the other hand, are
necessarily invariant under coordinate transformations. An important class of
such functions was introduced by Chern \cite{chern-curv-integra,
complex-manifolds-potential} in extending the notion of class functions
$f(g^{-1}xg)=f(x)$ from group theory. From (\ref{cartan-2}) follows that the
curvature form transforms under the adjoint representation of $\SO{n}$. Natural
choices are therefore the determinant and the trace of a polynomial in $\Omega$,
whose differential form is of the same order as the volume form of $\Sigma$ or
any submanifold thereof. Chern defined the Euler form or Euler class
\begin{equation}\label{euler-form}
\text{Pf}(\Omega) = \frac{1}{n!}\epsilon^{i_1\ldots i_n}\Omega_{i_1i_2}\wedge
\ldots \Omega_{i_{n-1}i_n}
\end{equation}
for even dimensional manifolds $n=2k$ and its integral as the Euler
characteristic 
\begin{equation}\label{euler-characteristic}
\chi(\Sigma) = \frac{(-1)^k(2k)!}{(4\pi)^k k!} \int_\Sigma \text{Pf}(\Omega)\;,
\end{equation}
with the normalization chosen such that its result is whole-numbered for the
sphere $\chi(S^2) = 2$ and the $g$-holed torus $\chi(T^2_g) = 2-2g$. The
integral is a topological invariant and central for many areas of mathematics
and physics \cite{nash}. It is therefore not surprising to discover that the
Euler form also enters the discussion of hard particle physics as the
intersection probability of particle stacks.  

The Euler class is the highest possible form for even dimensional manifolds
from which derives a series of invariant differential forms by successive
variation $\delta \omega = g^{-1}dg$. The resulting Chern-Simons classes
\cite{complex-manifolds-potential, alvarez-ginsparg} determine the failure of
the form to be invariant under the coordinate transformations
(\ref{transform-1}). As an example consider the case of $n=2$ dimensional
manifolds with the transition function $g= \exp{(i\lambda)} \in\text{U}(1)$. The
curvature reduces to the exterior derivative $\Omega = d\omega$ and its
variation $\delta \omega = id\lambda$ to a $i\mathbb{R}$ valued function:
\begin{equation}\label{gb-variation}
\delta \int_\Sigma \epsilon_{ij}\Omega_{ij} = \delta
\int_{\partial\Sigma} \epsilon_{ij} \omega_{ij} = \int_{\partial\Sigma}
i d\lambda = \int_{\partial\partial\Sigma} i\lambda\;.
\end{equation}
When the first integrant is rewritten by the Gaussian curvature $K$, the second
by the geodesic curvature $\kappa_g$ and the last integrant by the interior
angles, we obtain from (\ref{euler-characteristic}) the Gauss-Bonnet equation
for the 2-dimensional surface $\Sigma$
\begin{equation}\label{gauss-bonnet}
2\pi\chi(\Sigma) = \int_\Sigma K d\sigma + \int_{\partial\Sigma} \kappa_g ds + 
\sum_i (\pi - \alpha_i) 
\end{equation}
with non-contractible curves along $\partial \Sigma$ and additional vertices at
the singular points. To get a better understanding of the origin of these
additional contributions, remember that the Euler form counts the angular change
of the normal vector, while moving over the surface of the embedded manifold.
For smooth, Riemannian surfaces this is always $4\pi$, but boundaries and
singular points contribute additional angular changes and generate the
Chern-Simons terms.

It can be shown \cite{complex-manifolds-potential, alvarez-ginsparg} that the
two equations $\Omega = d\omega$, $\delta \omega = id\lambda$ generalize for the
Euler form $Q_{2k}^0 := \text{Pf}(\Omega)$ for arbitrary even dimension to a
sequence of characteristic classes
\begin{equation}
\delta Q_{2k}^0 = dQ_{2k-1}^0\;,\quad \delta Q_{2k-m}^{m-1} = dQ_{2k-m-1}^m
\end{equation}
for $m=1,\ldots, 2k$. Each variation now produces a new characteristic form of
one order less than its predecessor. And in the same way as the geodesic
curvature is an invariant form for the 1-dimensional curve $\partial \Sigma$, it
is natural to apply the odd differential forms of $Q_{2k-m-1}^m$ to odd
dimensional manifolds of non-trivial homotopy group. Euler's linking number and
the intersection number of hollow spheres are special cases of these forms. In
the notation of \cite{alvarez-ginsparg}, they correspond to $Q^1_2$ and $Q^2_3$
and derive from the Euler class of a 4 and 6 dimensional manifold. However, for
convex geometries, which we will consider in the following, it is not
necessary to take these classes into account.

Apart from the geometric interpretation of a Riemannian manifold, there is also
the relation to Lie groups, whose vielbein and connection forms constitute the
basis of a Lie algebra \cite{kobayashi, helgason, santalo-book} represented by
the matrix 
\begin{equation}\label{iso}
\sigma_A =
\begin{pmatrix}
\omega_{ij} & \theta_i \\
-\theta_j & 0
\end{pmatrix}
\in \text{iso}(n)\;,
\end{equation}
whose elements satisfy the Maurer-Cartan equations
\begin{equation}
\begin{split}
d\sigma^A &= \omega^A_{\;B}\wedge\sigma^B =
\frac{1}{2}C^A_{BC}\sigma^C\wedge\sigma^B\\
d^2\sigma^A &= 0 =
\frac{1}{2}C^A_{BC}C^C_{DE}\sigma^B\wedge\sigma^D\wedge\sigma^E\;.
\end{split}
\end{equation}
They are related by the inner derivation $i_B \sigma_A= \sigma_A(X_B)=
\eta_{AB}$ to the more commonly used commutation relation $[X_A,X_B] = C_{AB}^C
X_C$ and Jacobi identity. The corresponding Lie group is the Euclidean or
isometric group that locally splits into the semi-direct product $\text{ISO}(n)
=\SO{n} \ltimes E_n$ of rotations and translations. Its Lie algebra elements
$\omega_{ij}$ and $\theta_i$ transform under the mapping (\ref{transform-1}) and
span a $n(n+1)/2$ dimensional space consisting of the $n(n-1)/2$ connection and
$n$ vielbein forms.

The integral over all rotations and translations is therefore related to Haar's
measure of the isometric group
\begin{equation}\label{kin-measure}
\begin{split}
&\wedge_{i<j=1}^n \omega_{ij}\wedge_{i=1}^n\theta_i \\
&= \wedge_{i=1}^{n-1}\omega_{in}\wedge_{i=1}^{n-2}\omega_{i n-1}\wedge\ldots
\wedge \omega_{12}\wedge d\text{vol}(E_n) \\
&= dS^{n-1}\wedge dS^{n-2}\wedge\ldots \wedge dS^1\wedge d\text{vol}(E_n)\\
&= d\text{vol}(\SO{n})\wedge d\text{vol}(E_n)\;,
\end{split}
\end{equation}
where we made use of the coset representation:
\begin{equation}
\SO{k}/\SO{k-1} = S^{k-1}\;.
\end{equation}
Evaluating the integral yields then a product of volumes of spheres, with
values:
\begin{equation}\label{volume}
O_k = \text{vol}(S^k) = \frac{2\pi^{\frac{k+1}{2}}}{\Gamma(\frac{k+1}{2})}\;,
\end{equation}
whose first elements are $O_1 =2\pi$, $O_2= 4\pi$, \ldots .

\subsubsection{Extrinsic Geometry}
Up to now, we have only considered the intrinsic properties of the particles'
geometry. However, the movement in a background space requires the choice of a
suitable embedding. For physical reasons it is natural to consider the flat
Euclidean space and to imbed the $k$ dimensional particle, e.g., into the first
coordinate directions of the local frame $(e_1, \ldots, e_k, e_{k+1}, \ldots
e_n)$ with the corresponding nontrivial coordinate transformations
$\text{ISO}(n)/(\text{ISO}(k)\times\text{ISO}(n-k))$. To avoid the additional
problems that occur when discussing this complicated coset structure, we will
restrict the dimension of the embedding to $k=n-1$. The group $\text{ISO}(1)$
consists then only of the translation in one direction and can be explicitly
separated in the following equations. 

This choice is the simplest possible embedding and at the same time also the
physically most relevant one. The manifold $D \hookrightarrow\mathbb{R}^n$ is
now a $n$-dimensional domain in $\mathbb{R}^n$ and bounded by its surface
$\partial D$. Following the outline of \cite{chern-1}, we choose the outward
normal direction of the surface to point along $e_n$, such that the tangential
directions of $\partial D$ correspond to the first $n-1$ elements of the local
frame $(e_1,\ldots, e_{n-1}, e_n)$ of $\mathbb{R}^n$. The corresponding
directions are differenced by the index convention:
\begin{equation}\label{pfaff-index}
i,j = 1,\ldots n \quad \text{and} \quad \alpha,\beta = 1,\ldots, n-1\;.
\end{equation}

The associated Pfaff system \cite{kobayashi, bruhat} of the integrable
submanifold is then defined by the constraint
\begin{equation}\label{pfaff}
\theta_n = 0 \quad \text{on} \quad \partial D\;.
\end{equation}
Applied to the vanishing torsion of the Riemannian manifold
\begin{equation}\label{pre-rod}
0 = d\theta_n = \omega_{n \alpha}\wedge \theta_\alpha = h_{\alpha \beta}
\theta_\beta\wedge \theta_\alpha = \kappa_\alpha \nu_\alpha \wedge \nu_\alpha\;,
\end{equation}
it allows an algebraical solution of the equation by the symmetric matrix
$h_{\alpha\beta}$ and to define the principal curvatures $\kappa_\alpha$ and
principal vectors $\nu_\alpha$ as its eigenvalues and orthonormal eigenvectors.
In the form, $de_n= \kappa_\alpha \nu_\alpha e_\alpha$, it is also known as
Rodrigues formula.

Splitting the $n$-dimensional curvature $\Omega^{(n)}$ into normal and
tangential directions 
\begin{equation}
\begin{split}
\Omega^{(n)}_{n\alpha} &= d\omega_{n\alpha}-\omega_{\alpha\beta}\wedge
\omega_{\beta n}\\
& = D^{(n-1)} \omega_{n\alpha}\\
\Omega^{(n)}_{\alpha\beta} &= d\omega_{\alpha\beta} -
\omega_{\alpha\gamma}\wedge\omega_{\gamma\beta}- \omega_{\alpha n}\wedge
\omega_{n\beta}\\
&= \Omega_{\alpha\beta}^{(n-1)} - \omega_{\alpha n}\wedge \omega_{n \beta}
\end{split}
\end{equation}
yields the Gauss and Gauss-Codazzi equations \cite{kobayashi}, further
reducing to
\begin{equation}\label{gauss}
\Omega_{\alpha\beta}^{(n-1)} = \omega_{\alpha n}\wedge \omega _{n\alpha}
\quad \text{and} \quad 
D^{(n-1)} \omega_{n\alpha} = 0
\end{equation}
in the case of flat embedding. The first equation relates the intrinsic
curvature of the particle to the normal connection forms of the embedding.
Whereas the vanishing of $\omega_{n\alpha}$ under the $n-1$ dimensional
covariant derivative ensures the decoupling of the normal coordinate
transformations from the tangential ones; the forms $\omega_{n\alpha}$ are
therefore horizontal \cite{kobayashi}, without the need of introducing
equivariant differential forms \cite{greub}. 

In the definition of the embedding we have assumed that the normal vector $e_n$
points outward from the compact particle surface. This corresponds to a special
gauge choice in the $\text{O}(n)$ coordinate transformations of $\mathbb{R}^n$
and restricts the group to $\SO{n}$. But this local gauge does not extend
globally, where both orientations $\mathbb{Z}_2 \cong \text{O}(n)/\SO{n}$ have
to be taken into account. The Euler characteristic, derived by the intrinsic
curvature (\ref{cartan-2}) and by Gauss's equation (\ref{gauss}), will
correspondingly differ by a factor of two
\begin{equation}\label{euler-imbed}
\chi(\partial D) = 2\chi(D \hookrightarrow \mathbb{R}^n)\;.
\end{equation}

The kinematic measure (\ref{kin-measure}) of an embedded particle of odd
dimension $n=2k+1$ can now be calculated by combining (\ref{euler-form},
\ref{euler-characteristic}, \ref{gauss}, \ref{euler-imbed}) and observing
that the normalization of the Euler characteristic is proportional to
$O_{2k}=2(4\pi)^k k!/(2k)!$, as follows from (\ref{volume})
\begin{align}\label{int-m}\nonumber
&\int \wedge_{i<j=1}^{2k+1}\omega_{ij}\wedge_{i=1}^{2k+1}\theta_i\\
&= \int \text{Pf}(\Omega)\wedge d\text{vol}(\SO{2k})\wedge
d\text{vol}(E_{2k+1})\\[0.5em] 
&= \chi(\partial D)\,\text{vol}(\SO{2k+1})\,\text{vol}(E_{2k+1})\;.\nonumber
\end{align}
For a 3-dimensional manifold in $\mathbb{R}^3$, the corresponding integral
reduces to 
\begin{align}\label{int-3}\nonumber
\int & \omega_{12}\omega_{13}\omega_{23}\theta_1\theta_2\theta_3
= \int \kappa_1\kappa_2 \nu_1\nu_2 \cdot \int \omega_{12}\,
d\text{vol}(\mathbb{R}^3)\\[0.5em]\nonumber 
&= \int K(\partial D) dS_A \cdot 2\pi\,V\\
&= 8\pi^2 V\int \frac{1}{4\pi} K(\partial D) \delta(\vec{n}\vec{r}_A)
d^3r_A\\[0.4em]
&= 8\pi^2\,V\,\chi(\partial D)\nonumber
\end{align}
with the volumes of $\text{vol}(\mathbb{R}^3)=V$ and $\text{vol}(\SO{3}) =
8\pi^2$. 

Note that the kinematic measure of a Riemannian manifold would vanish for
dimensional reasons, as the vielbein and connection forms are not independent.
It is therefore necessary first to interpret the integrant as the Haar's
measure and only afterwards to incorporate the geometric constraints. 

This equation is of course closely related to Chern's original derivation of the
Euler class \cite{chern-curv-integra}. Here however, the difference lies
in the relation between geometry and isometric group, which focuses on the
alternative interpretation as the kinematic measure of a particle, moving
in a flat background. For two intersecting particles it thus determines the
intersection probability, averaged over all rotations and translations. It is
therefore identical to the second virial integral and explains the appearance of
the Gauss-Bonnet equation (\ref{gauss-bonnet}) in the calculations of Isihara
and Kihara \cite{kihara-1}, Rosenfeld \cite{rosenfeld2}, and Wertheim
\cite{wertheim-1}. 
\subsection{The One, Two, and Three Particle Intersections}
\label{subsec:branching}
\subsubsection{Comments on Integral Geometry}
The generalization of (\ref{int-m}) to two and more intersecting particles leads
us into the field of integral geometry, whose differential geometric formulation
goes back to Minkowski \cite{minkowski}, Weyl \cite{weyl-tube}, Blaschke
\cite{blaschke}, Santalo \cite{santalo-book}, and Chern \cite{chern-1, chern-2,
chern-3}, who observed that the invariant forms of integral geometry can be
traced back to the Euler class (\ref{int-m}). One intriguing result is the
fundamental kinematic equation \cite{santalo-book} 
\begin{equation}\label{fke}
\begin{split}
\frac{V_n}{\gamma_n} \int\limits_{\text{ISO}(n)} \chi(D_1\cap g D_2)dg &=
\sum_{k=0}^n \binom{n}{k}M_k(D_1)M_{n-k}(D_2)\\
\gamma_n= \text{vol}(\SO{n})\; &, \quad V_n = \frac{1}{n}O_{n-1}
\end{split}
\end{equation}
and the observation that the coupled geometry of two intersecting manifolds
reduces to a simple pairwise product of Minkowski measures or integrals of mean
curvature $M_k$. For $n=3$ it reproduces the equation of Isihara and Kihara of
the second virial coefficient. Actually, they used for their calculation an
early result of Minkowski \cite{minkowski}. In fact, it was the starting
point for our current investigation and offers a direct, albeit less general,
approach of deriving the intersecting probability, which is why we have added
their calculation in a somewhat clarified form in appendix \ref{sub:appendix_a}.

There are several ways to derive the fundamental kinematic equation (\ref{fke}).
Probably the simplest one uses the expansion of the Steiner polynomial
\cite{santalo-book}, another one Blaschke's cut and paste construction
\cite{blaschke} of subspaces. The most fundamental, although more elaborate
approach is Chern's explicit derivation \cite{chern-1} of the Euler class from
the kinematic measure (\ref{int-m}). Its advantage is the explicit local
formulation in connection forms that will be important for its decoupling into
Rosenfeld's weight functions. This ansatz is therefore the natural starting
point for relating Rosenfeld's approach to integral geometry.

The generalization of (\ref{int-m}) to a particle stack $\St_{k+1}$ is easily
achieved but requires some normalization to get a well defined result. First, we
have to fix the position and orientation of one particle in $\St_{k+1}$ to
remove the volume dependence on the embedding space $V=\text{vol}
(\mathbb{R}^3)$, generated by moving the stack in the background manifold.
Furthermore, it is useful to define the kinematic measures of the particle
domain $D$ and its surface $\Sigma$
\begin{equation}\label{norm_measure}
\begin{split}
dD &= \omega_{12}\omega_{13}\omega_{23}\wedge \theta_1 \theta_2 \theta_3\\
d\Sigma &= \omega_{12}\wedge \theta_1 \theta_2
\end{split}
\end{equation}
analogously to (\ref{def-measure}). The kinematic measure of (\ref{int-m}) or
(\ref{int-3}) generalizes then to the integral average of $k+1$ particles
\begin{equation}\label{stack-inter}
\begin{split}
\frac{1}{8\pi^2} & \frac{1}{V}\int_{St_{k+1}}\,dD_1\wedge\ldots \wedge
dD_{k+1}\\
&= \frac{1}{4\pi}\int_{\St_{k+1}} K(\partial
\St_{k+1})\,\delta(\vec{n}\vec{r}_A) \, d^3r_A \\
& \qquad\qquad\qquad \times d\gamma_2\wedge \ldots \wedge d\gamma_{k+1}\;,
\end{split}
\end{equation}
with the Gaussian curvature $K$ integrated over the domain $A = \St_{k+1}$
at fixed kinematic measure $\gamma$, as defined in (\ref{def-measure}), and
integrated over the center of gravity, represented by $\gamma_1$.

The boundary of the stack $\partial \St_{k+1}$ can be determined by the
algebraic relations of the homology operator \cite{bott-tu}. As an example,
consider two intersecting manifolds that itself have no boundary $\partial^2
D=0$. The application of $\partial$ to the second order stack
\begin{equation}\label{boundary}
\partial (D_1\cap D_2) = \partial D_1 \cap D_2 + D_1\cap \partial D_2 +
\partial D_1\cap \partial D_2\;.
\end{equation}
is thus a sum of intersections, wherein each successive application of
$\partial$ reduces the dimension by one. This restricts the possible number of
boundary operations to the dimension of the embedding space by the constraint
$\partial^4X = 0$ for any 3-dimensional manifold $X$. The infinite number of
virial contributions, shown in FIG.~\ref{fig:zero-loop}, reduces therefore to
the derivation of three Euler forms, corresponding to one, two, and three
particles. 

The calculation of (\ref{stack-inter}) can be further simplified by including
the physical constraint of indistinguishable particles. To obtain the correct
combinatorial pre-factors, let us define the formal sum
\begin{equation}\label{devisor}
\hat{n} = \sum_{i=1}^M D_i \,\rho_i
\end{equation}
of 1-particle densities and domains. It is the homologous operator of
Rosenfeld's weight densities and parallels the notion of a divisor in algebraic
geometry. The representation of the free-energy functional in $\hat{n}$ reduces
the problem of determining the boundary of the stack of $k+1$ different
particles $\partial \St_{k+1}$ to the corresponding analysis of a stack of $k+1$
identical manifolds, whose boundary reduces to a sum of three terms
\begin{equation}\label{boundary-stack}
\begin{split}
\partial \St_{k+1} & = (k+1)\,\Sigma\cap\St_k +
(k+1)k\,\Sigma\cap\Sigma\cap\St_{k-1}\\
& + (k+1)k(k-1)\,\Sigma\cap\Sigma\cap\Sigma\cap\St_{k-2}
\end{split}
\end{equation}
in the shorthand notation $\Sigma = \partial D$. Using the linearity of the
Euler form and its vanishing for odd dimensional manifolds, it translates to
the corresponding Gaussian curvature
\begin{equation}\label{boundary-2}
\begin{split}
K(\partial \St_{k+1}) &= (k+1)\, K(\Sigma)+ (k+1)k\, K(\Sigma\cap \Sigma)\\
& + (k+1)k(k-1)\, K(\Sigma \cap\Sigma \cap \Sigma)
\end{split}
\end{equation}
that will be derived in the following. The first two terms are known from Chern
\cite{chern-1}, who obtained the result for two intersecting manifolds of
arbitrary dimension. An independent approach was used by Wertheim
\cite{wertheim-1}. However, the three-particle intersection is new and will be
presented parallel to the summary of the previous two cases. The corresponding
generalization of Chern's approach to an arbitrary number of particles and
dimensions has been developed in \cite{korden} and will now be applied to three
dimensions. 
\subsubsection{The One Particle Euler Form}
Let us begin with the simplest case $K(\Sigma)$ of one particle, moving in a
background of $k$ domains. Following the derivation of (\ref{int-3}) the
product of the connection forms can be rewritten in the principal basis
$\omega_{13}\wedge\omega_{23}= \kappa_1\kappa_2\nu_1\wedge\nu_2$, reducing
the kinematic measure of $\Sigma$
\begin{equation}
\begin{split}
&\frac{1}{8\pi^2}\frac{1}{V}\int_{\Gamma(D_1)} dD_1 =
\frac{1}{8\pi^2}\frac{1}{V}\int_{\Gamma(D_1)}
\omega_{13}\omega_{23}\omega_{12}\theta_1\theta_2\theta_3\\
&= \frac{1}{4\pi}\int_{A=D_1} \kappa_G\, \delta(\vec{n}\vec{r}_A)\,d^3r_A
\end{split}
\end{equation}
with the Gaussian curvature $\kappa_G$ and a factor of $2\pi$ from the integral
over $\omega_{12}$. The first part of the integral (\ref{boundary-stack}) for a
stack can now be written as
\begin{align}\label{1-part-int}\nonumber
&\frac{1}{4\pi}\int_{\substack{\Gamma(D_2\times \ldots \times D_{k+1})\\
\times \St_{k+1}}}
K(\Sigma_1, \vec r_A)\,\delta(\vec{n}\vec{r}_A) \,
d^3r_A\,dD_2 \ldots dD_{k+1}\\\nonumber
&=\int\frac{1}{4\pi} \kappa_G(\partial D_1, \vec r_A)\,\delta(\vec{n}\vec{r}_A)
\,
\Theta(D_2, \vec r_A)\ldots \Theta(D_{k+1}, \vec r_A)\\\nonumber
&\qquad\qquad\qquad  \times d^3r_A\,d\gamma_2 \ldots d\gamma_{k+1}\\
&= \int \omega_\chi^{(1)} \omega_v^{(2)}\ldots
\omega_v^{(k+1)} d^3r_A\,d\gamma_2\ldots d\gamma_{k+1}
\end{align}
in the weight functions 
\begin{equation}\label{weight-1}
\omega_\chi^{(i)} = \frac{1}{4\pi}\kappa_G\, \delta(\partial D_i)\;,
\quad \omega_v^{(i)} = \Theta(D_i)\;,
\end{equation}
where the integration domain has been formally extended to the complete
embedding space $V$ by the Dirac- and Heaviside-function $\delta$ and
$\Theta$, with $\delta(\partial D_i)$ understood as restricting the volume
integration to the surface $\delta(\vec{n}\vec{r}_A)$ at the intersection
point $\vec r_A \in \St_k$.
\subsubsection{The Two Particle Euler Form}
Some more efforts requires the derivation of the second Euler form
$K(\Sigma_1\cap\Sigma_2)$ that determines the angular change between the
two normal vectors at the 1-dimensional intersection submanifold. It 
parallels the geodesic curvature $\kappa_g$ of the Gauss-Bonnet formula
(\ref{gauss-bonnet}) and can be seen as the real space generalization of the
Chern-Simons class. Its derivation begins with the construction of a proper
coordinate system at the intersection space. Let us introduce the bases
$\Sigma_1:(e_1, e_2^{(1)}, e_3^{(1)})$ and $\Sigma_2:(e_1, e_2^{(2)},
e_3^{(2)})$ with the common direction $\Sigma_1\cap\Sigma_2:e_1$ along the
1-dimensional submanifold and the intersection angle 
\begin{equation}\label{angle}
\cos{(\phi_{12})}=(e_3^{(1)} e_3^{(2)})\quad \text{for}\quad  
0 \leq \phi_{12} < 2\pi\;.
\end{equation}
Following \cite{korden}, we define the intersection determinant
\begin{equation}\label{intersection_matrix}
M_k=\det{(e_3^{(i)} e_3^{(j)})}|_{i,j=1}^k
\end{equation}
for $k$ intersecting surfaces. The first two cases are:
\begin{equation}\label{intersect_2_3}
\begin{split}
M_2 &= 1- c_{12}^2 = s_{12}^2 \\
M_3 &= 1- c_{12}^2 - c_{13}^2 - c_{23}^2 + 2c_{12}c_{13}c_{23}\;,
\end{split}
\end{equation}
where we used the shorthand notation:
\begin{equation}
\begin{split}
s_{ij}:=\sin{(\phi_{ij})}\;&,\quad c_{ij}:=\cos{(\phi_{ij})}\\
s(\gamma) := \sin{(\gamma)}\;&,\quad c(\gamma):=\cos{(\gamma)}\;.
\end{split}
\end{equation}

The local frame of the intersection manifold in $\mathbb{R}^3$ is spanned by the
vector field $e_1, e_3^{(1)}, e_3^{(2)}$ for $\phi_{12}\neq 0$, from which one
obtains an orthonormal basis by the Gram-Schmidt process
\begin{align}\label{s-basis}\nonumber
v_3 & = e_3^{(1)}\\
v_2 & = \frac{1}{s_{12}} \bigl( e_3^{(2)} - c_{12}\,e_3^{(1)}
\bigr)\\
v_1 & = e_1\;.\nonumber
\end{align}
As explained before, the Euler characteristic counts the angular change of the
normal vector, while moving from $e_3^{(1)}$ to $e_3^{(2)}$. To interpolate
between those two vectors, we introduce a $\SO{2}$ rotation in the range $0\leq
\gamma \leq \phi_{12}$ 
\begin{equation}
\begin{split}
\eta_3 &= c(\gamma)v_3 + s(\gamma)v_2\\ 
\eta_2 &= -s(\gamma)v_3 + c(\gamma)v_2\\
\eta_1 &= e_1\;.
\end{split}
\end{equation}
One of the two equivalent vectors, $\eta_3$ or $\eta_2$, is now the new
outward pointing normal direction. Let us chose $\eta_3$ and derive the
corresponding Euler density for the intersection $\Sigma_1\cap \Sigma_2$:
\begin{equation}
\begin{split}
\eta_1d\eta_3 & \wedge\eta_2d\eta_3 \\
&= \frac{1}{s_{12}} \left[s(\phi_{12}-\gamma)\omega_{13}^{(1)}+ s(\gamma)
\omega_{13}^{(2)}\right] \wedge d\gamma\;,
\end{split}
\end{equation}
with the definition $\omega_{13}^{(i)}=e_1de_3^{(i)}$ of the new connection
forms for the particles $i=1,2$. Integrating over $\gamma$ 
\begin{equation}\label{euler-2}
\begin{split}
\int \eta_1d\eta_3\wedge\eta_2d\eta_3 &= \frac{1-c_{12}}{s_{12}}
[\omega_{13}^{(1)} + \omega_{13}^{(2)}]\\
&= K(\Sigma_1\cap \Sigma_2) d(\Sigma_1\cap
\Sigma_2)
\end{split}
\end{equation}
yields the differential Euler form. Observe, that the angular dependent factor
$(1-c_{12})/s_{12}$ remains finite even in the limit of anti-parallel
vectors $\phi_{12}\to \pi$ when the remaining kinematic measure is included,
which will be derived below.

At the intersection $\Sigma_1\cap \Sigma_2$, a $\SO{2}$ transformation
(\ref{transform-1}) relates the vector frames of the two particles
\begin{equation}\label{rot-2}
\begin{split}
e_3^{(2)} &= c_{12} e_3^{(1)} + s_{12} e_2^{(2)}\\
e_2^{(2)} &= -s_{12} e_3^{(1)} + c_{12} e_2^{(2)}\\
e_1^{(2)} &= e_1^{(1)}
\end{split}
\end{equation}
and the boundary condition $\theta_3^{(i)}|_{\partial \Sigma_i}=0$ their
corresponding Pfaffian systems (\ref{pfaff}). The transformed differential
forms 
\begin{equation}\label{forms-trafo}
\begin{split}
\theta_3^{(2)} & = c_{12} \theta_3^{(1)} + s_{12} \theta_2^{(1)}\\
\omega_{13}^{(2)} & = c_{12}\omega_{13}^{(1)} + s_{12}\omega_{12}^{(1)}\\
\omega_{23}^{(2)} & = \omega_{23}^{(1)} + d\phi_{12}
\end{split}
\end{equation}
are therefore understood modulo $\theta_3^{(1)}, \omega_{\alpha 3}^{(1)}$. With
these relations, the reduced kinematic measure of $\Sigma_1\cap\Sigma_2$ can be
derived, with the first particle fixed in the embedding space and the second
one free to move:
\begin{align}\label{kin-meas-2}\nonumber
d( & \Sigma_1 \cap \Sigma_2)\wedge dD_2\\\nonumber
& = \theta_1^{(1)}\wedge
\theta_1^{(2)}\theta_2^{(2)}\theta_3^{(2)}\omega_{12}^{(2)}
\omega_{13}^{(2)}\omega_{23}^{(2)} \\
& = (s_{12})^2\;
\theta_1^{(1)}\theta_2^{(1)}\omega_{12}^{(1)}\wedge 
\theta_1^{(2)}\theta_2^{(2)}\omega_{12}^{(2)}\wedge d\phi_{12}\\\nonumber
& = (s_{12})^2\; d\Sigma_1\wedge d\Sigma_2\wedge d\phi_{12}\;,\nonumber
\end{align}
with the kinematic measure of the surface defined in (\ref{norm_measure}). The
decoupling of the Euler form (\ref{euler-2}) and the kinematic measure
(\ref{kin-meas-2}) for two intersecting particles is a central property of
integral geometry \cite{santalo-book} and follows from the $\text{ISO}(3)$
invariance.

Next, we transform $(e_1^{(i)},e_2^{(i)},e_3^{(i)})$ into the orthonormal
coordinate system of the principal frame $(\vec \nu_1^{(i)}, \vec \nu_2^{(i)},
\vec n^{(i)})$, changing the notation for the normal direction $\vec n=e_3$ to
be consistent with Rosenfeld's and Wertheim's convention. The 3-dimensional
cross product of the normal vectors
\begin{equation}\label{tangential}
e_1 = v_1 = v_2\wedge v_3 = \frac{1}{s_{12}}\;\vec{n}^{(2)}\times
\vec{n}^{(1)}\;
\end{equation}
points now into the tangential direction of the intersection. Combining the
Euler form and the kinematic measure, we obtain the intersection probability
between two particles:
\begin{equation}\label{chern-pre-trans}
\begin{split}
& \frac{1}{8\pi^2}\frac{1}{V} \int_{\Gamma(D_1\times D_2)}dD_1\wedge dD_2 \\
& = \frac{1}{4\pi} \int_{\Gamma(D_2)}\int_{\Sigma_1\cap \Sigma_2}
K(\Sigma_1\cap \Sigma_2) d(\Sigma_1\cap \Sigma_2)dD_2\\
& = \frac{1}{4\pi} \int_{\Gamma(D_2)}\int_{\Sigma_1\cap \Sigma_2}
\frac{1-c_{12}}{s_{12}}
[\omega_{13}^{(1)} + \omega_{13}^{(2)}]\;dD_2
\end{split}
\end{equation}
integrated over the intersection volume $A = D_1\cap D_2$ and the
kinematic measure with $\phi_{12}\in \Gamma(D_2)$.

The transformation of the connection forms from the old reference system 
to the principal frame was done by Chern \cite{chern-1}. However, Wertheim's
tensorial representation \cite{wertheim-1} (see also \cite{rosenfeld2,
rosenfeld-gauss2, mecke-fmt, goos-mecke}) has the advantage to be more closely 
related to Rosenfeld's definition of weight functions. In order to keep the
discussion self-contained, we have included Wertheim's derivation in appendix
\ref{sub:appendix_b} and present here only the result. 

Using the diagonal form of the Euclidean metric $\mathbb{I}$ and the curvature
tensor $\mathbb{K}$ 
\begin{equation}\label{metric-1}
\begin{split}
\mathbb{I} & = \vec \nu_1 \otimes \vec \nu_1 +  \vec \nu_2 \otimes \vec \nu_2 
+ \vec n \otimes \vec n\\
\mathbb{K} & = \kappa_1\, \vec \nu_1 \otimes \vec \nu_1 + \kappa_2\, \vec \nu_2
\otimes \vec \nu_2\;,
\end{split}
\end{equation}
Rodrigues formula (\ref{pre-rod}) yields the form:
\begin{equation}
e_1 d e_3 = e_1 \mathbb{K} e_1\, ds 
= e_1 [\bar\kappa(\mathbb{I}-\vec n \otimes \vec n)
 + \Delta] e_1\,ds
\end{equation}
with the mean and tangential curvature
\begin{equation}\label{many-curvatures}
\bar{\kappa} = \frac{1}{2}(\kappa_1+\kappa_2)\;,\quad 
\Delta  =\frac{1}{2}(\kappa_1-\kappa_2)(\nu_1\otimes \nu_1 
- \nu_2\otimes \nu_2)\;.
\end{equation}
With this change of notations and appendix \ref{sub:appendix_b}, we finally
obtain Wertheim's representation of the kinematic measure
\begin{align}\label{wertheim-2}\nonumber
&\frac{1}{4\pi} \int_{\Gamma(D_2)}\int_{\Sigma_1\cap \Sigma_2}
K(\Sigma_1\cap \Sigma_2) d(\Sigma_1\cap \Sigma_2)dD_2\\
& = \frac{1}{4\pi} \int_{\Gamma(D_2)} \int_{A= D_1\cap D_2} \Bigl[
(\mathbb{I}-\vec{n}^{(1)}\otimes \vec{n}^{(2)})(\bar\kappa^{(1)} +
\bar\kappa^{(2)}) \Bigr.\\ 
&\quad  \Bigl.-\frac{\vec{n}^{(1)}\Delta^{(2)}\vec{n}^{(1)} +
\vec{n}^{(2)}\Delta^{(1)}\vec{n}^{(2)}}{1+\vec{n}^{(1)}\vec{n}^{(2)}}
\Bigr]\nonumber\\[0.75em]
& \quad \times  \delta(\vec{n}^{(2)}\vec{r}_A)
\delta(\vec{n}^{(1)}\vec{r}_A)\, d^3r_A \,dD_2\nonumber
\end{align}
integrated over $\vec{r}_A\in D_1\cap D_2$ and $\Gamma(D_2)$.

Now it is a simple task to expand the denominator in the geometric series 
\begin{equation}
\frac{1}{1+\vec{n}^{(1)}\vec{n}^{(2)}} = 1-\vec{n}^{(1)}\vec{n}^{(2))} 
+ (\vec{n}^{(1)})^{\otimes 2}(\vec{n}^{(2))})^{\otimes 2} \pm \ldots
\end{equation}
of tensor products and to rewrite the integral in the weight functions
\begin{equation}\label{2-part-int}
\begin{split}
& \frac{1}{4\pi} \int_{\substack{\Gamma(D_2\times \ldots \times D_{k+1})\\
\times \Sigma_1\cap \Sigma_2}}
K(\Sigma_1\cap\Sigma_2)
d(\Sigma_1\cap \Sigma_2) dD_2\ldots dD_{k+1}\\
&= \int_{\substack{\Gamma(D_2\times \ldots \times D_{k+1})\\
\times \St_{k+1}}}
\Bigl[ \omega_{\kappa 0}^{(1)}\omega_{\sigma 0}^{(2)}
-\omega_{\kappa 1}^{(1)}\omega_{\sigma 1}^{(2)} 
- \sum_{L=0}^\infty \omega_{\Delta L+2}^{(1)}\omega_{\sigma L}^{(2)} \Bigr.\\
& \qquad \Bigl.+ (1\leftrightarrow 2)\;\Bigr]\,\omega_v^{(3)}\ldots
\omega_v^{(k+1)} \; d^3r_A\, d\gamma_2\ldots d\gamma_{k+1}
\end{split}
\end{equation}
with the extended basis set of Rosenfeld's weight functions:
\begin{equation}\label{all-weights}
\begin{split}
\omega_\chi(D) & = \frac{1}{4\pi}\kappa_G\,\delta(\partial D)\\
\omega_{\kappa L}(D) & = \frac{1}{4\pi} \bar{\kappa} (n)^{\otimes L}\,
\delta(\partial D) \\
\omega_{\Delta L}(D) & = \frac{1}{4\pi}\Delta (n)^{\otimes L}\,
\delta(\partial D)\\[0.25em]
\omega_{\sigma L}(D) & = (n)^{\otimes L}\,\delta(\partial D)\\[0.75em]
\omega_v(D) & = \Theta(D)\;.
\end{split}
\end{equation}
with the abbreviation:
\begin{equation}
\delta(\partial D) = \delta(\vec{n}\vec{r}, \partial D)\;.
\end{equation}
The normalization of the curvature dependent terms has been chosen to absorb
the overall constant of $4\pi$. In the following we will see that these are all
basis functions for 3-dimensional, convex particles.
\subsubsection{The Three Particle Euler Form}
The third and last case is the Euler form for three intersecting particles. Its
intersection $\Sigma_1\cap \Sigma_2\cap \Sigma_3$ consists of points, whose
corresponding Euler class is a 0-form and independent of $\omega_{ij}$. It
therefore parallels the angular dependent part of the Gauss-Bonnet equation
(\ref{gauss-bonnet}). 

As before (\ref{s-basis}), the three normal vectors $e_3^{(1)}, e_3^{(2)},
e_3^{(3)}$ are converted into an orthonormal basis by the Gram-Schmidt method:
\begin{align}\label{schmidt}\nonumber
v_1 & = e_3^{(1)}\\
v_2 & = \frac{1}{\sqrt{M_2}}\bigl( e_3^{(2)} - (e_3^{(2)} v_1)\,v_1
\bigr)\\\nonumber
v_3 & = \frac{M_2}{\sqrt{M_3}}\bigl( e_3^{(3)} - (e_3^{(3)} v_2)\,v_2 -
(e_3^{(3)} v_1)\,v_1 \bigr)
\end{align}
and extended to the local frame
\begin{equation}
\eta_i = R_{ij}(\gamma_1,\gamma_2,\gamma_3)v_j\;,\quad R_{ij}\in \SO{3}\;,
\end{equation}
interpolating between the three normal directions. Here, we can use the
same argument that let to the simplification of (\ref{kin-measure}) and replace
the product of the connection forms by the volume of $\SO{3}$ in Euler angles:
\begin{equation}\label{3-inter}
K(\Sigma_1\cap\Sigma_2\cap \Sigma_3) = \int \sin{(\gamma_2)}d\gamma_1 d\gamma_2
d\gamma_3\;.
\end{equation}
However, $\gamma_2$ measures the angle between the vector and the $x_2$-axis and
not the angle between the normal vectors. We therefore introduce a new
coordinate system
\begin{equation}\label{new-coord}
\begin{split}
e_3^{(1)} = &\begin{pmatrix}0\\0\\1\end{pmatrix}\;,\;
e_3^{(2)} = \begin{pmatrix}0 \\ s(\alpha_1) \\ c(\alpha_1) \end{pmatrix}\;, \\
e_3^{(3)} &= \begin{pmatrix} s(\alpha_3)s(\alpha_2) \\ s(\alpha_3)c(\alpha_2) \\
c(\alpha_3) \end{pmatrix}
\end{split}
\end{equation}
that is related to the Euler angles by 
\begin{equation}\label{euler-trafo}
\begin{split}
c(\gamma_2) = s(\alpha_1)c(\alpha_2) & s(\alpha_3) + c(\alpha_1)c(\alpha_3)\\
c(\gamma_1) = c(\alpha_1)\;, & \;\; c(\gamma_3) = c(\alpha_3)\;.
\end{split}
\end{equation}
The new representation of the Euler form (\ref{3-inter})
\begin{equation}\label{3-intersection}
\begin{split}
K( & \Sigma_1 \cap\Sigma_2\cap \Sigma_3) \\
& = \int \sin{(\alpha_1)} \sin{(\alpha_2)} \sin{(\alpha_3)}
d\alpha_1d\alpha_2d\alpha_3\\
&= (1-\cos{(\phi_{12})})(1-\cos{(\phi_{13})})(1-\cos{(\phi_{23})})\\[0.6em]
&= (1-\vec{n}^{(1)} \vec{n}^{(2)})(1-\vec{n}^{(1)}
\vec{n}^{(3)})(1-\vec{n}^{(2)} \vec{n}^{(3)})
\end{split}
\end{equation}
is a symmetric polynomial in the normal vectors. The remaining integration over
the intersection space $\Sigma_1\cap\Sigma_2\cap \Sigma_3$ reduces to a finite
sum over its intersection points
\begin{align}\label{3-euler-form}\nonumber
&\int K(\Sigma_1 \cap\Sigma_2\cap \Sigma_3)d(\Sigma_1\cap\Sigma_2\cap\Sigma_3)
\\
& = \frac{1}{2}\int_{\St_3} (1-c_{12})(1-c_{13})(1-c_{23}) \\
&\qquad\quad  \times 
\delta(\vec{n}^{(1)}\vec{r}_A) \delta(\vec{n}^{(3)}\vec{r}_A)
\delta(\vec{n}^{(3)}\vec{r}_A)\, d^3r_A\,\nonumber\\[0.4em]
&= \frac{1}{2}\sum_{\text{pt}\in \Sigma_1\cap \Sigma_2\cap \Sigma_3}
\frac{1-c_{12}}{s_{12}}
\frac{1-c_{13}}{s_{13}}\frac{1-c_{23}}{s_{23}}\;,\nonumber
\end{align}
where relation (\ref{d-3}) and the vector basis (\ref{new-coord}) for the
normal directions has been used
\begin{equation}
|\vec{n}^{(1)}(\vec{n}^{(2)}\times \vec{n}^{(3)})| = s_{12} s_{13} s_{23}\;.
\end{equation}
Furthermore, a factor $1/2$ has been added to compensate for the double covering
of the integration range, when instead of the Euler angles $0\leq \phi, \psi <
2\pi$ and $0\leq \theta < \pi$ the symmetric choice of the intersection angles
\begin{equation}\label{range}
0\leq \phi_{12},\phi_{13},\phi_{23} < 2\pi\;.
\end{equation}
is used. 

Next, we have to determine the kinematic measure with one of the three
particles fixed in space. The derivation parallels that of (\ref{kin-meas-2})
and begins with the coordinate transformation of $dD_2\wedge dD_3$. Following
the approach of \cite{korden}, we rotate the locale frame of particle $D_3$ by
the matrix
\begin{equation}
R_1(\gamma_1)=
\begin{pmatrix}
1 & 0 & 0 \\
0 & c_1 & s_1 \\
0 & -s_1 & c_1
\end{pmatrix}\;,\;
\end{equation}
in the $3\to 1$ direction and derive the new vielbein and connection forms for
$D_1$:
\begin{equation}
\begin{split}
\omega_{13}^{(3)} & = c_1\omega_{13}^{(1)} - s_1\omega_{12}^{(1)}\\
\omega_{23}^{(3)} & = \omega_{23}^{(1)} - d\gamma_1 \\
\theta_3 ^{(3)} & = -s_1 \theta_2^{(1)} + c_1  \theta_3^{(1)}
\end{split}
\end{equation}

The same calculation has to be done for particle $D_2$, where the matrix
\begin{equation}
R_{23}(\gamma_2,\gamma_3)=
\begin{pmatrix}
c_2 & s_2c_3 & s_2s_3 \\
-s_2 & c_2c_3 & c_2s_3 \\
0 & -s_3 & c_3
\end{pmatrix}
\end{equation}
generates a $2\to 1$ rotation
\begin{equation}
\begin{split}
\omega_{23}^{(2)} &= s_2s_3\omega_{12}^{(1)} - s_2c_3\omega_{13}^{(1)} +
c_2\omega_{23}^{(1)} -c_2 d\gamma_3 \\
\omega_{12}^{(2)} &= c_3\omega_{12}^{(1)} + s_3 \omega_{13}^{(1)} - d\gamma_2\\
\theta_2^{(2)} &= -s_2\theta_1^{(1)} + c_2c_3\theta_2^{(1)} + c_2s_3
\theta_3^{(1)}\;.
\end{split}
\end{equation}

The forms in the normal direction of $D_1$ vanish by the constraint
(\ref{pfaff}). We can therefore set the corresponding terms of $\theta_3^{(1)},
\omega_{13}^{(1)}$, and $\omega_{23}^{(1)}$ to zero and insert the transformed
elements into $dD_2\wedge dD_3$. Performing an additional coordinate shift
$\gamma_2 \to \gamma_2 +\pi/2$ and the change of basis (\ref{euler-trafo}) to
transform from the Euler into the intersection angles, we finally obtain the
reduced kinematic measure
\begin{equation}
\begin{split}
& dD_2 \wedge dD_3\\
&= (s_{12}s_{13}s_{23})^2 d\Sigma_1 d\Sigma_2 d\Sigma_3
d\phi_{12}d\phi_{13}d\phi_{23} 
\end{split}
\end{equation}
with the kinematic measure of the surface $d\Sigma$ defined in
(\ref{norm_measure}).

Collecting terms, the Euler form (\ref{3-euler-form}) intersecting with $k-2$
further particles is determined by:
\begin{align}\label{kin-meas-3}
&\frac{1}{4\pi}\int_{\Gamma(D_2\times \ldots \times D_{k+1})}
\sum_{\{\text{pt}\}}
K(\Sigma_1 \cap\Sigma_2\cap \Sigma_3) dD_2 \ldots dD_{k+1}\nonumber\\
&=\frac{1}{8\pi} \int_{\substack{\Gamma(D_2\times \ldots \times D_{k+1})\\
\times \St_{k+1}}}\nonumber\\[0.4em]
&\times
(1-\vec{n}^{(1)}\vec{n}^{(2)})\,(1-\vec{n}^{(1)}\vec{n}^{(3)})\,(1-\vec{n}^{(2)}
n^{(3) })\\[0.4em]
&\times \delta(\vec{n}^{(1)}\vec{r}_A) \delta(\vec{n}^{(2)}\vec{r}_A)
\delta(\vec{n}^{(3)}\vec{r}_A)\, d^3r_A d\gamma_2 \ldots d\gamma_{k+1}\;,
\nonumber 
\end{align}
and can be rewritten in the basis of the weight functions, defined in
(\ref{all-weights}), after expanding the product of (\ref{kin-meas-3}):
\begin{equation}\label{3-euler}
\begin{split}
&\frac{1}{4\pi} \int_{\Gamma(D_2\times \ldots \times D_{k+1})}
\sum_{\{\text{pt}\}} 
K(\Sigma_1\cap\Sigma_2\cap \Sigma_3) dD_2\ldots
dD_{k+1} \\
&=  \frac{1}{8\pi}\int_{\substack{\Gamma(D_2\times \ldots \times D_{k+1})\\
\times \St_{k+1}}}
\Bigl[
\omega_{\sigma 0}^{(1)}\omega_{\sigma 0}^{(2)}\omega_{\sigma 0}^{(3)} \\
&-  \omega_{\sigma 0}^{(1)}\omega_{\sigma 1}^{(2)}\omega_{\sigma 1}^{(3)}
-   \omega_{\sigma 0}^{(2)}\omega_{\sigma 1}^{(1)}\omega_{\sigma 1}^{(3)}
-\omega_{\sigma 0}^{(3)}\omega_{\sigma 1}^{(1)}\omega_{\sigma_1}^{(2)}\\[0.5em]
&+  \omega_{\sigma 2}^{(1)}\omega_{\sigma 1}^{(2)}\omega_{\sigma 1}^{(3)}
+   \omega_{\sigma 2}^{(2)}\omega_{\sigma 1}^{(1)}\omega_{\sigma 1}^{(3)}
+\omega_{\sigma 2}^{(3)}\omega_{\sigma 1}^{(1)}\omega_{\sigma 1}^{(2)}\\[0.5em]
&-  \omega_{\sigma 2}^{(1)}\omega_{\sigma 2}^{(2)}\omega_{\sigma 2}^{(3)}
\Bigr]\, \omega_v^{(4)}\ldots \omega_v^{(k+1)}\, d^3r_A\,d\gamma_2 \ldots
d\gamma_{k+1}\;.
\end{split}
\end{equation}
As required, the result is invariant under cyclic permutations of the indices
$(1,2,3)$. 

For the first two integrals (\ref{2-part-int}, \ref{1-part-int}) it was
possible to scale the pre-factor to one by a suitable definition of the weight
functions. The same is not possible for (\ref{3-euler}), as it depends only on
the previously defined weights. The three particle integral has therefore an
overall pre-factor of $1/8\pi$.

The three intersection probabilities (\ref{2-part-int}, \ref{1-part-int},
\ref{3-euler}) are complicated polynomials in the weight functions. However,
here we have shown, by explicit calculation, that these three cases are all we
have to consider under the given restrictions on the manifolds. The five
different types of weight functions (\ref{all-weights}) are complete in this
sense and provides the basis for higher loop orders. The grouping of the weight
functions into five classes can be stated more formally by their scaling
dimension under the coordinate transformation $\vec{r}\to \lambda \vec{r}$. 

Let us summarize the results of this section:
\begin{theorem}\label{theorem}
The Euler form $\omega_\chi$ of the kinematic measure of a stack $\St_k$ of
3-dimensional, convex Riemannian manifolds decomposes into a symmetric sum of
weight functions 
\begin{align}\label{branching}\nonumber
&\omega_\chi(\Sigma_1 \cap \ldots \cap \Sigma_k) 
 = C_{A_1 \ldots A_k}\,\omega_{A_1}(\Sigma_1)\,\ldots 
\omega_{A_k}(\Sigma_k)\\[0.5em]
&\omega_\chi(\Sigma_1\cap\ldots\cap D_{k-1}\cap D_k)\\
&\qquad\qquad \qquad \qquad  =
\omega_\chi(\Sigma_1\cap\ldots\cap D_{k-1})\omega_v(D_k)\nonumber \\[0.5em]
&\hspace{5em} \omega_\chi(\Sigma\cap D) = \;\omega_\chi(\Sigma)
\omega_v(D)\;,\nonumber
\end{align}
where an implicit summation over the multi-index $A\in \{\chi, v, \kappa L,
\Delta L\}$ for $L=0,1,2,\ldots$ is understood. The numerical values of
the coefficients $C_{A_1 A_2 \ldots}$ follow from (\ref{2-part-int},
\ref{1-part-int}, \ref{3-euler}). They depend on the dimension of the embedding
space and the particle but are otherwise independent of the manifold's
geometry. 

The weight functions (\ref{all-weights}) provide a complete basis set, in which
the intersection integrals can be expanded. They are unique with respect to the
Euler form. Their scaling dimensions group the weight functions into four
subclasses:
\begin{equation}\label{scaling}
[\omega^i_\chi] = 3\;,\; [\omega^i_{\kappa L}] = 
[\omega^i_{\Delta L}] = 2\;,\; [\omega^i_{\sigma L}] = 1\;,\; [\omega^i_v] = 0
\end{equation}
\end{theorem}
\section{Resummation and the Rosenfeld Functional}\label{sec:rosenfeld}
\subsection{The Functional of Rosenfeld and Tarazona}\label{subsec:ros-tara}
\subsubsection{Rosenfeld's Three Postulates}
The local decomposition of the kinematic formula for one, two, and three
particle intersections clarifies the mathematical aspects of Rosenfeld's
approach. However, it remains to combine the resulting weight functions into the
free-energy functional. A first naive attempt of inserting the reduced
virial integrals into the corresponding expansion of the chemical potential
\begin{equation}\label{thermodyn}
\begin{split}
\beta \mu &= \beta \mu_{\text{id}} + \sum_{n=1}^\infty \beta_n \rho^n \\
\beta\mathcal{F} &= \int \mu(\vec{r}) d\rho(\vec{r})d^3r
\end{split}
\end{equation}
fails. The reason lies in the decoupling of the particle density $\rho$ from
its geometric properties $\omega_A$ that allows to add a particle by the
integration of (\ref{thermodyn}) without adding the particle's volume
$\omega_v$. To find a corresponding generalization, let us reconsider
Rosenfeld's derivation of the functional \cite{rosenfeld-freezing} (see also
\cite{mcdonald}).

The infinite number of weight functions (\ref{all-weights}) reduces to a finite
subset for spheres, whose principal curvatures $\kappa_1=\kappa_2$ causes the
$\Delta$-dependent terms to vanish. The second virial integral of a mixture of
hard spheres with $M$ components reduces therefore to a finite sum of only six
weight functions.
\begin{equation}\label{m-convolute}
\begin{split}
-f_{ij}(|\vec{r}_i-\vec{r}_j|) &= \sum_{A_1,A_2} C_{A_1 A_2}\,
\omega_{A_1}^i\otimes \omega_{A_2}^j \\
& = \omega_\chi^i\otimes \omega_v^j + \omega_{\kappa 0}^i \otimes 
\omega_{\sigma 0}^j - \omega_{\kappa 1}^i\otimes
\omega_{\sigma 1}^j\\[0.5em]
& + (i\leftrightarrow j)
\end{split}
\end{equation}
where $i,j=1,\ldots, M$ runs over all types of spheres. The tensor product is a
short form of the convolute integral
\begin{equation}\label{product}
\omega_{A_1}^i\otimes \omega_{A_2}^j = 
\int_{D_i\cap D_j} \omega_{A_1}^i(\vec{r}_A-\vec{r}_i)\,
\omega_{A_2}^j(\vec{r}_A-\vec{r}_j)\, d^3r_A
\end{equation}
depending on the particle positions $\vec{r}_i, \vec{r}_j$ in the
embedding space $\mathbb{R}^3$ and the intersection point $\vec{r}_A \in D_i\cap
D_j$. From the decoupling of the integral measure (\ref{m-convolute}) into
single particle contributions follows the splitting of the entire second virial
integral, weighted by the 1-particle densities $\rho_i(\vec{r}_i)$:
\begin{equation}\label{ros-second-vir}
\begin{split}
&-\frac{1}{2}\beta_1(D_i,D_j)\\
&= \sum_{A_1,A_2,i,j}C^{A_1 A_2}\int\rho(i)\rho(j)
(\omega_{A_1}^i \otimes \omega_{A_2}^j) d\gamma_i d\gamma_j d^3r_A\\
&=\sum_{A_1,A_2}C^{A_1 A_2} \int_{D_i\cap D_j} n_{A_1}(\vec{r}_A)\,
n_{A_2}(\vec{r}_A)\, d^3r_A
\end{split}
\end{equation}
written in the weight densities:
\begin{equation}\label{weight-density}
n_A(\vec{r}_A) = \sum_{i=1}^M\int_{\Gamma(D_i)} \rho_i(\vec{r}_i)\,
\omega^i_A(\vec{r}_A - \vec{r}_i)\, d\gamma_i\;.
\end{equation}

As has been discussed \ref{subsec:stacks}, the pairing of one weight function
with the 1-particle density is a consequence of the single intersection domain
of the second virial cluster. However, it is natural to generalize this
construction further to particles with $k$ intersection centers. The
corresponding integral then combines $k$ weight functions with the 1-particle
density:
\begin{equation}\label{k-point}
\begin{split}
n_{A_1,\ldots, A_k} &(\vec{r}_{A_1},\ldots,\vec{r}_{A_k}) \\
&= \sum_{i=1}^M \int_{\Gamma(D_i)} \rho_i(\vec{r}_i) \prod_{\nu=1}^k
\omega_{A_\nu}^i(\vec{r}_{A_\nu} - \vec{r}_i)\, d\gamma_i\,
\end{split}
\end{equation}
generalizing the 2-point densities of the exact third virial integral
(\ref{3-exact}). Such ``$k$-point densities'' are the central objects in
analyzing higher loop diagrams. With increasing loop order increases also the
order of the $k$-point densities. This can be seen by assuming that all $g$
loops begin and end at the same particle. The loop diagrams then decouple into
sets of $k$-point densities for $2\le k \leq 2g$. The only diagrams that contain
1-point densities are therefore the intersection stacks of $g=0$, as has been
explained \ref{subsec:stacks}.

From the observation that the leading contribution of the free-energy factorizes
into products of weight densities, Rosenfeld postulates three assumptions about
the structure of the functional: Firstly, the free-energy is an analytic
function in the weight densities, i.e. it allows a polynomial expansion in $n_A$
\begin{equation}\label{free-funct}
\beta\mathcal{F}^{\text{ex}}([n_A]) = 
\int \Phi^{\text{ex}}_{\text{R}}([n_A])\,d^3r\;.
\end{equation}
Of course, we have seen in section \ref{subsec:stacks} that this assumption is
not true in general. However, the functional form of $\mathcal{F}^{\text{ex}}$
can be further restricted by observing that the integral (\ref{free-funct}) has
to be invariant under coordinate scaling. The second assumption is therefore
that the free-energy functional is a homogenous polynomial under the
transformation $\vec{r}\to \lambda^{-1} \vec{r}$ with the scaling dimension
\begin{equation}\label{scaling-dimension}
[\Phi^{\text{ex}}_{\text{R}}] = - [d^3r] = 3
\end{equation}
of the free-energy. The possible combinations of weight functions are therefore
constrained by their scaling dimensions (\ref{scaling}) with the exception of
the scale independent $\omega_v$:
\begin{equation}
\begin{split}
\Phi^{\text{ex}}_{\text{R}}([n_A]) & = f_1(n_v)n_\chi + 
f_2(n_v)n_{\kappa 0} n_{\sigma 0} + 
f_3(n_v) n_{\kappa 1} n_{\sigma 1}\\
& + f_4(n_v) n_{\sigma 0}^3 + f_5(n_v)n_{\sigma 0} n_{\sigma 1} n_{\sigma 1}\;.
\end{split}
\end{equation}
With the third postulate, Rosenfeld further assumes that the functional is a
solution of the scaled particle differential equation \cite{rosenfeld-structure,
mcdonald}. In this way it is possible to determine the dependence of the unknown
functions $f_1, \ldots, f_5$ on the scale-invariant weight density $n_v$. The
free-energy functional is then known up to the integration constants of the
solutions of the differential equation. For $f_1, f_2, f_3$, they can be read
off from the second virial contribution; but the constants for $f_4$ and $f_5$
have to be determined by comparison with analytical results obtained by
alternative methods. The functional has thus the preliminary form
\cite{rosenfeld2}:
\begin{equation}\label{ros_prelim}
\begin{split}
\Phi^{\text{ex}}_{\text{prelim}}([n_\alpha])& = -n_\chi\ln{(1-n_v)} +
\frac{n_{\kappa 0}n_{\sigma 0} - n_{\kappa 1} n_{\sigma 1}}{1-n_v}\\
&+ \frac{1}{24\pi} \frac{n_{\sigma 0}^3 -3 n_{\sigma 0}
n_{\sigma 1} n_{\sigma 1}}{(1-n_v)^2}\;.
\end{split}
\end{equation}
Later on, it has been shown that this functional leads to an unphysical
singularity, when the positions of the spheres were constrained to lower
dimensions \cite{crossover-ros-1, tarazona-rosenfeld}. The source for the
occurring divergence is the third term in the functional. This led Rosenfeld and
Tarazona to look for alternative third order polynomials compensating the
singularity. Several suggestions were made \cite{tarazona-rosenfeld,
crossover-rosenfeld-2, tarazona} and compared to simulations. The most promising
modification today is Tarazona's \cite{tarazona} replacement:
\begin{equation}\label{3-order-correction}
\begin{split}
\Phi_3 & = \frac{1}{16\pi}\left[ \prod_{(ij)}(1-e_3^{(i)} e_3^{(j)})
-[e_3^{(1)},e_3^{(2)},e_3^{(3)}]^2\right] \\ 
& = \frac{1}{16\pi} \left[ (1-c_{12})(1-c_{13})(1-c_{23}) - M_3\right]\\
\end{split}
\end{equation}
with $M_3$ from (\ref{intersect_2_3}). Comparing this semi-heuristic result to
equation (\ref{3-intersection}), identifies the first term as the three-particle
intersection probability of the stack. In \cite{tarazona-rosenfeld,
comparision-ros} it has been shown that the corresponding correction of the
functional (\ref{ros_prelim}) by this term alone is in excellent agreement with
simulation data of the bulk-fluid free-energy of hard spheres. The fluid phase
is therefore well described by the intersection probability of stacks.
However, it has been shown in \cite{tarazona-rosenfeld} that the Lindemann ratio
for the fcc-lattice is underestimated by this functional. This is corrected by
the second part of (\ref{3-order-correction}), improving the equation of state
for the solid region \cite{tarazona}. In the next section we will argue that
this term is part of the 1-loop correction of the third virial diagram.

The final form of the Rosenfeld functional for hard spheres \cite{tarazona} is
obtained by replacing the third term of (\ref{ros_prelim}) by Tarazona's
expression (\ref{3-order-correction}):
\begin{equation}\label{ros_func}
\begin{split}
&\Phi^{\text{ex}}_{\text{R}}([n_\alpha]) = -n_\chi\ln{(1-n_v)} +
\frac{n_{\kappa 0}n_{\sigma 0} - n_{\kappa 1}n_{\sigma 1}}{1-n_v}
- \frac{3}{16\pi}\\
& \times \frac{n_{\sigma 0}n_{\sigma 1}n_{\sigma 1}
-n_{\sigma 1} n_{\sigma 2}n_{\sigma 1} 
+ n_{\sigma 2}n_{\sigma 2}n_{\sigma 2} - n_{\sigma 0}
n_{\sigma 2}n_{\sigma 2}}{(1-n_v)^2}\\
\end{split}
\end{equation}
This result provides one of the currently best approximations of the fluid phase
structure of hard spheres, only surpassed by the White Bear version
\cite{white-bear-1, white-bear-2}. However, this improvement has been obtained
by adjusting the functional to simulation data, whereas the correction
(\ref{3-order-correction}) is geometrically motivated. Apart from the $M_3$-term
in (\ref{3-order-correction}), we have already derived all of its contributions
and pre-factors from the 0-loop order.
\subsubsection{Replacing the Scaled Particle Differential Equation}
The chemical potential enters the fundamental measure theory via the scaled
particle differential equation. Its origin is a semi-heuristic relation between
the chemical potential and the pressure $\mu^{\text{ex}}_i \to pv_i$ in the low
density limit that becomes exact at diverging particle volume $v_i \to \infty$.
This limit allows to relate the chemical potential of the free-energy
$\mathcal{F}$ to the pressure representation of the grand potential $-pV =
\Omega = \mathcal{F} - \rho_i \delta \mathcal{F} / \delta \rho_i$. Introducing
the functional derivative:
\begin{equation}\label{variation}
\frac{\delta \rho_i(\vec{r}_i)}{\delta \rho_j(\vec{r}_j)} =
\delta_{ij}\,\delta(\vec{r}_i-\vec{r}_j)\;,
\end{equation}
which selects the weight function when applied to a weight density
\begin{equation}
\begin{split}
\frac{\delta}{\delta \rho_j(\vec{r}_j)} n_A(\vec{r}_A) 
&= \int \sum_{i} \omega^i_A(\vec{r}_A-\vec{r}_i)
\delta_{ij}\,\delta(\vec{r}_i-\vec{r}_j) d^3r_i\\
&= \omega^j_A(\vec{r}_A-\vec{r}_j)\;,
\end{split}
\end{equation}
the chemical potential $\mu^{\text{ex}}_i$ of the free-energy functional has
the form:
\begin{align}\label{spt-chem}\nonumber
\beta\mu^{\text{ex}}_i & (\vec{r}, \vec{r}_i) =
\frac{\delta \Phi^{\text{ex}}_{\text{R}}(\vec{r})}{\delta \rho_i(\vec{r}_i)} 
= \sum_A \frac{\partial \Phi^{\text{ex}}_{\text{R}}}{\partial
n_A} \frac{\delta n_A(\vec{r})}{\delta \rho_i(\vec{r}_i)}\\\nonumber
&=\frac{\partial \Phi^{\text{ex}}_{\text{R}}}{\partial n_v} 
\omega^i_v(\vec{r},\vec{r}_i) +
\sum_{A\neq v} \frac{\partial \Phi^{\text{ex}}_{\text{R}}}{\partial
n_A} \omega^i_A(\vec{r},\vec{r}_i) \\\nonumber
& \hspace{-0.75em} \underset{v_i\to \infty}{=} (\beta\, p^{\text{ex}} +
\rho)\omega^i_v(\vec{r},\vec{r}_i) \\
& = \Bigl(-\Phi^{\text{ex}}_{\text{R}} + 
\sum_A n_A \frac{\partial \Phi^{\text{ex}}_{\text{R}}}{\partial
n_A} + n_\chi \Bigr)\omega_v^i(\vec{r},\vec{r}_i)\;,
\end{align}
assuming that all contributions of $\omega_A^i$ vanish in the $v_i\to\infty$
limit except for $\omega_v^i$. From this follows the scaled particle
differential equation:
\begin{equation}\label{spt}
\Phi^{\text{ex}}_{\text{R}} + \frac{\partial
\Phi^{\text{ex}}_{\text{R}}}{\partial n_v} - \sum_A
n_A \frac{\partial \Phi^{\text{ex}}_{\text{R}}}{\partial n_A} =
n_\chi\;.
\end{equation}

The arguments leading to this result are by no means trivial: The scaled
particle limit allows the identification of the particle volume $v_i$ as the
embedding volume $V$, resulting in the unpaired index $v$ in the last two lines
of (\ref{spt-chem}). Another striking feature is the dependence of the chemical
potential on the two different coordinate systems of the particles $\vec{r}_i\in
D_i$ and those of the intersection region $\vec{r}\in \St_k$. This
indicates a further difficulty in identifying the chemical potential as an
external potential coupled to the particle density. To obtain a symmetric
formulation in the densities $\rho_i$ and $n_A$, let us define the chemical
potential for the particle volume $n_v$:
\begin{equation}\label{v-chem}
\Psi_v(\vec{r}) := \beta \frac{\delta \mathcal{F}^{\text{ex}}([n_A])}{\delta
n_v(\vec{r})}\;.
\end{equation}

In principle it is possible to define an infinite set of chemical potentials for
the weight functions $\omega_A^i$. However, $\Psi_v$ is the only physically
relevant one. This can be realized in two different ways: Firstly, $\delta n_v$
is again scale invariant, which follows from $[\rho_i] = -[d^3r] = 3$ and
$[\omega_v] = 0$. $\Psi_v$ has therefore the same scale dependence as the
free-energy. This complies with the interpretation as the energy change by
inserting a particle into the system and the observation that $\omega_v^i$ is
the only scale invariant weight function. Secondly, it follows from
(\ref{boundary-2}) that the intersection probability of a stack $\St_k$ of order
$k>3$ will only change by a factor $\omega^i_v$, when an additional particle is
inserted. This corresponds to a formal integration over $\omega^i_v$ coupled to
the particle density $\rho_i$. 

The functional derivative (\ref{v-chem}) can be inverted by integration
\begin{equation}\label{ros-rel-mu}
\begin{split}
\beta\mathcal{F^{\text{ex}}} &= \int \Psi_v(\vec{r}) \delta n_v(\vec{r})
:= \int \Psi_v dn_v d^3r\\
&  = \int \Phi^{\text{ex}}(\vec{r}) d^3r
\end{split}
\end{equation}
and relates the chemical potential to Rosenfeld's free-energy density. It also
allows a natural interpretation of $\Psi_v$ as the integral of the functional
derivative 
\begin{equation}\label{chem-func}
\mu_{iv}(\vec{r}_i, \vec{r}) = \frac{\delta}{\delta \rho_i(\vec{r}_i)}\frac{
\mathcal{F}^{\text{ex}}([n_A])}{\delta n_v(\vec{r})}\;.
\end{equation}
The two derivatives with respect to $\rho_i$ and $n_A=\rho_i \omega^i_{A}$ are
of course not independent from each other and do not commute $\mu_{iv}\neq
\mu_{vi}$. It is therefore important not to interchange the order in the
integration
\begin{equation}\label{mu-rel}
\mathcal{F}^{\text{ex}} = \int \Psi_v \delta n_v = \int (\mu_{iv}\,
\delta \rho_i) \delta n_v\;.
\end{equation}

Now, $\mu_{iv}$ has the right structure for generalizing the virial expansion
(\ref{thermodyn}) to the weight function depending terms $\beta_n(\omega^i_A)
\rho_i^n$. Furthermore, it is extensible to arbitrary loop orders. Inserting the
expansion (\ref{thermodyn}) into (\ref{mu-rel}) with subsequent integration over
$\rho_i$ gives a general relation between the virial expansion and the
free-energy density (\ref{ros-rel-mu}):
\begin{equation}\label{virial-fmt}
\Phi^{\text{ex}}([n_A], \vec{r}) = cn_v + \sum_{k=1}^\infty \frac{1}{k+1}\int
\rho^{k+1} \beta_k dn_v\;.
\end{equation}
The integration constant $c$ is itself a functional of the remaining weight
densities $n_A$ for $A\neq v$ to be determined by comparing $\Phi^{\text{ex}}$
to the low-density limit. However, the scaling dimension restricts the possible
dependence to $c\propto n_\chi$, with a universal constant to be determined in
the next section.

Equation (\ref{virial-fmt}) generalizes the virial expansion (\ref{thermodyn})
of the free-energy to the functional form depending on the weight densities. It
is an exact relation and independent of the semi-heuristic scaled particle
theory. Once the virial coefficients are known, we can derive the functional by
a simple integration over $n_v$ for any loop order.
\subsection{The 0-Loop Order of the Free-Energy Functional}
\label{subsec:ros_functional}
With the derivation of the intersection probability of particle stacks 
(\ref{branching}) and the virial expansion of the free-energy in terms
of the weight densities (\ref{virial-fmt}), we can finally put the pieces
together and prove our hypothesis (\ref{aim}) that Rosenfeld's functional
$\Phi_R^{\text{ex}}$ corresponds to the leading order $\Phi_0$ of the loop
expansion (\ref{top-exp}). This is done in two steps: deriving the virial
integrals for any diagram of zero order, and then adding them up into a
generating function.

In section \ref{subsec:stacks} we have seen that a Mayer cluster of loop order
$g$ decomposes into a series of topological diagrams
\begin{equation}\label{beta-exp}
\beta_k = \sum_{n=0}^g  \beta_k^n\;,
\end{equation}
of which the leading order $\beta_k^0$ corresponds to the intersection
probability of a stack $\St_{k+1}$. Following the discussion from section
\ref{subsec:branching}, the corresponding cluster integral
\begin{align}\label{virial-0}
& \beta_k^0 = \frac{1}{V} \frac{\sigma}{k!} \int_{\Gamma(D_1\times \ldots \times
D_{k+1})} f_{1,2} \ldots f_{k,k+1}\;
d\gamma_1\ldots d\gamma_{k+1}\nonumber \\
&= \frac{1}{k!} \int_{\substack{\Gamma(D_2\times \ldots \times D_{k+1})\\
\times \St_{k+1}}} K(\partial \St_{k+1})
d^3r_A\, d\gamma_2\ldots d\gamma_{k+1}
\end{align}
is identical to the averaged Euler form, integrated over the kinematic measure
of $k+1$ particles. Here we have used that the symmetry coefficient is $\sigma =
1$ and that the volume factor $V$ cancels after integrating over the coordinates
of the center of gravity. In principle it is possible to extend the integral to
mixtures of particles by including an additional index. However, this is not
necessary, as the final result will depend on the weight densities
(\ref{weight-density}), which automatically include the right combinatorial
factors. We can therefore restrict the discussion to a single class of particles
without loss of generality. 

The boundary of a stack of identical, 3-dimensional particles has been derived
in (\ref{boundary-stack}) and reduces to the sum of three contributions. The
branching rules of (\ref{branching}) can then be used to algebraically split
the Euler form of (\ref{virial-0}) into the volume dependent weight functions
\begin{equation}\label{b-split}
\begin{split}
\omega_\chi( & \partial \St_{k+1})= (k+1)\, \omega_\chi(\Sigma)\omega_v^k \\
& + k(k+1)\,\omega_\chi(\Sigma\cap\Sigma)\omega_v^{k-1}\\
&+ k(k+1)(k-1)\,\omega_\chi(\Sigma\cap\Sigma\cap\Sigma)\omega_v^{k-2}
\end{split}
\end{equation}
and further into the decoupled product of weight densities:
\begin{equation}\label{m-split}
\begin{split}
\omega_{\chi}(\Sigma\cap\Sigma\cap\Sigma) &=
C_{A_1 A_2 A_3}\omega_{A_1} \omega_{A_2} \omega_{A_3}\\
\omega_{\chi}(\Sigma\cap\Sigma) &= C_{A_1 A_2}\,
\omega_{A_1}\omega_{A_2}\\
\omega_{\chi}(\Sigma\cap D) & = C_{\chi v}\,\omega_\chi \omega_v
\end{split}
\end{equation}
where an implicit sum over the paired indices is understood. We also introduced
the trivial constant $C_{\chi v}=1$ to keep the notation symmetrical. In
anticipation of the following derivation of the Rosenfeld functional
(\ref{ros_func}), it is useful to separate the dependence on the highest and
lowest weight functions $\omega_\chi, \omega_v$ from the Euler form and to
introduce the index notation
\begin{equation}
A = (\chi, v, \alpha) = (\chi, v, \kappa L, \Delta L) 
\end{equation}
deduced from Theorem~\ref{theorem}.  

Inserting (\ref{b-split}) and (\ref{m-split}) into (\ref{virial-0}) yields
the virial integral for a stack
\begin{equation}\label{crude-virial}
\begin{split}
&\beta_k^0 = (k+1) \int [C_{\chi v}\; \omega_{\chi}\omega_v^k +
k\, C_{\alpha_1 \alpha_2} \omega_{\alpha_1} \omega_{\alpha_2}
\omega_v^{k-1} \\
& +k(k-1)\, C_{\alpha_1\alpha_2\alpha_3}\;
\omega_{\alpha_1} \omega_{\alpha_2} \omega_{\alpha_3}\,
\omega_v^{k-2}]\;d^3r_A \prod_{i=2}^{k+1} d\gamma_i
\end{split}
\end{equation}
of $k+1$ indistinguishable particles. The virial coefficient is a homogeneous
polynomial of order $k+1$ in the weight functions and combines with the
particle density $\rho^{k+1}$ to a polynomial of weight densities. Inserted
into (\ref{virial-fmt}), we obtain the result:
\begin{equation}\label{chemical}
\begin{split}
& \Psi_v^0([n_A]) = c+\sum_{k=1}\frac{1}{k+1}\rho^{k+1}\beta_k^0\\
& = c +  C_{\chi v}\,n_\chi\,[\frac{1}{1-n_v} - 1]
+ C_{\alpha_1 \alpha_2} \frac{n_{\alpha_1}n_{\alpha_2}}{(1-n_v)^2} \\
&+ 2\,C_{\alpha_1 \alpha_2 \alpha_3}
\frac{n_{\alpha_1}n_{\alpha_2}n_{\alpha_3}}{(1-n_v)^3}\;.
\end{split}
\end{equation}
The integration constant $c$ can now be uniquely determined by comparing it to
the ideal gas limit, where the $n_v$ dependence has to vanish. Inserting the
value $c = C_{\chi v}n_\chi$ and integrating over the $n_v$ density gives the
final excess free-energy functional of the 0-loop order:
\begin{equation}\label{fin-func1}
\begin{split}
\Phi^{\text{ex}}_0([n_A]) &= \int \Psi_v^0 dn_v\\
&=  -C_{\chi v}\; n_\chi \ln{(1-n_v)}
+ C_{\alpha_1 \alpha_2} \frac{n_{\alpha_1}n_{\alpha_2}}{1-n_v}\\
& + C_{\alpha_1 \alpha_2 \alpha_3}
\frac{n_{\alpha_1}n_{\alpha_2}n_{\alpha_3}}{(1-n_v)^2}\;.
\end{split}
\end{equation}
Comparing this result to the Rosenfeld functional (\ref{ros_prelim}), we have
finally proved our hypothesis (\ref{aim}). 

This result also allows a formal extension to $D$-dimensional particles embedded
into the odd dimensional $\mathbb{R}^D$. Because the Mayer expansion is
independent of the dimension of the physical system, nothing will change by this
generalization. Extending the boundary stack (\ref{boundary-stack}) to $D$
dimensions and the corresponding splitting of the Euler form (\ref{branching})
results in a free-energy functional
\begin{equation}\label{gen-func}
\begin{split}
\Phi^{\text{ex}}_0([n_\alpha]) & = \sum_{k=1}^D
C_{\alpha_1 \ldots \alpha_k} n_{\alpha_1}\ldots n_{\alpha_k}
\frac{\partial^k}{\partial n_v^k} \phi(n_v)\\
\phi(n_v) & = (1-n_v)\ln{(1-n_v)} + n_v
\end{split}
\end{equation}
that can conveniently be written by the generating functional $\phi$. The same
observation has been made before in \cite{crossover-rosenfeld-2}, where
$\phi(n_v)$ has been derived in the freezing limit, when the particles are
located in caverns. Here, we can see that the generating functional carries the
volume dependent parts of the boundary of the universal stack as defined in
(\ref{stack}). The Rosenfeld functional has now the simple interpretation as the
intersection probability of $\USt$.

Thus we have shown that the 0-loop order of the virial expansion leads to the
Rosenfeld functional. However, it only reproduces the first term of Tarazona's
correction (\ref{3-order-correction}). Therefore, one might guess that the
$M_3$-dependent part belongs to the 1-loop correction of the third virial order
(\ref{3-exact}) as will be investigated in a subsequent article.
\section{Discussion and Conclusion}\label{sec:conclusion}
In this article it has been shown that the Euler form $K(\partial \St_k)$
determines the intersection probability of a particle stack of order $k$ and
that its generating function reproduces Rosenfeld's functional. These results
explain and generalize Rosenfeld's previously unproven observation
\cite{rosenfeld-structure, rosenfeld2, rosenfeld-gauss2} that
the second virial integrand is related to the Gauss-Bonnet equation. For two
intersecting convex particles the results of Wertheim \cite{wertheim-1} and
Hansen-Goos and Mecke \cite{mecke-fmt, goos-mecke} are confirmed by explicitely
deriving the Euler form from first principles. However, going beyond the second
virial, we further derived the previously unknown Euler forms for $k\geq 3$ and
their splitting into weight functions. 

Motivated by the success of Rosenfeld's functional for the liquid region, we
made the Euler form the foundation of the fundamental measure theory and its
extension beyond the currently known functional. It has been shown that the
Mayer clusters of hard particles split into intersection diagrams that can be
classified by their number of loops and intersection points, where the latter
corresponds to a particle stack. The leading contribution, the 0-loop order, is
then the only part of the free-energy that can be represented by a functional
with only one intersection point.

From this follows that the fundamental measure theory allows the systematic
derivation of the free-energy functional for each loop order; a result that is
in fundamental contrast to DFT in quantum mechanics, where the development of a
functional is only restricted by the existence theorem of Hohenberg and Kohn
\cite{gross}. This property of hard particle physics is probably a consequence
of the invariance of the Euler form under geometric deformations. As long as the
homotopy type and therefore the topology does not change, we obtain the same
functional form. And even if we include complex geometries like tori or hollow
spheres, the additional terms still derive from an Euler form. The only
constraints we have to consider are of physical nature and are related to
concave geometries.

The infinite number $L=0,1,\ldots$ of tensorial weight functions provide a
practical problem in the calculation of higher loop orders. Since we cannot
derive an infinite set of integrals, it is necessary to stop at a certain order.
A first hint gives Wertheim's calculation of the third virial integral for
prolate and oblate spheroids \cite{wertheim-3, wertheim-4}. He shows that the
aspect ratio $\lambda \leq 10$ differs from the simulated result by less than
3\%, when the $L\leq 2$ terms are included. This indicates that the expansion
of the denominator $1+\vec{n}^{(1)}\vec{n}^{(2)}$ is fast converging for most of
the physically interesting cases. 

Also of importance is the influence of the number of loops and intersection
points. As explained in section \ref{subsec:stacks}, each intersection point of
a diagram is dressed by the universal stack, as shown in
FIG.~\ref{fig:zero-loop}, whose free-energy contribution is already known from
the 0-loop order. Consequently, each intersection carries a factor of
$(1-n_v)^{-1}$ and $(1-n_v)^{-2}$. From this follows that the divergence of
the resummed third virial integral $\Phi_{(1,3)}$ of
FIG.~\ref{fig:third-virial-reg} is at least of order $(1-n_v)^{-3}$. The
influence of diagrams decreases therefore significantly with their number of
intersection points. We therefore expect no new physical effects by including
higher intersection orders. This is consistent with our hypothesis that only
higher loop orders correspond to long range effects between particles, as
indicated by the generating function of all 1-loop diagrams.

Another aspect worth considering is the dimensional influence of the particles
and their embedding space. If the codimension is larger than 1, the particles do
not necessarily intersect, while approaching each other. The mathematical
formulation is then more complicated and requires the introduction of
equivariant differential forms \cite{greub}; in the physical literature this is
known from BRST quantization \cite{zinn-justin}. We have also seen that
the Euler form vanishes for odd dimensions and gets replaced by higher order
invariant forms. This is a consequence of the Bott periodicity \cite{spin} and
offers a direct link between the mathematical and physical properties. It is
even possible that this relation can be further extended to a more detailed
understanding of the relation between topology, geometry and the physical phase
structure of particles. For example, one might ask, if the geometry of a
particle and its mixtures can be tested by their phase diagrams?

An important step in this direction is the numerical calculation of weight
functions and the minimization of the grand potential functional
\cite{mcdonald}. For the 3-dimensional particles it is possible to reduce the
problem to a triangulation of the surface and to replace the connection form by
a sum over the outward angles, analogously to the derivation of the Gauss-Bonnet
equation. The resulting polyhedrons are then placed into a Voronoi diagram,
whose boundaries are varied until the minimum of the free-energy has been
obtained. This approach would allow the analysis of even more complicated
particle distributions than the isotropic or periodic structures
investigated so far. In addition, it would also allow a better understanding of
the origin of phase transitions. For instance, the particles in the nematic and
smectic phase are parallel oriented, minimizing the 0-loop contribution of
the free-energy by setting one or more of the intersection angles to zero.
However, understanding such effects requires the derivation of higher loop
orders and will therefore be postponed to the next article.
\section{Acknowledgment}
Professor Matthias Schmidt is kindly acknowledged for stimulating discussions
and valuable comments on the manuscript.
This work was performed as part of the Cluster of Excellence "Tailor-Made Fuels
from Biomasse", which is funded by the Excellence Initiative by the German
federal and state governments to promote science and research at German
universities.
\appendix
\section{}\label{sub:appendix_a}
It is enlightening to compare the local formulation of Chern \cite{chern-1} to
the approach of Minkowski \cite{minkowski}, which was the basis for the
calculation of Isihara and Kihara \cite{isihara-orig, kihara-1}. We will
therefore give a short summary of their derivation that led to the first general
equation of the second virial coefficient of convex particles. Let $p_i\in D_i$
be the coordinate vector of the two convex particles $i=1,2$. The excluded
volume under translation and rotation of the particles is then calculated by
first deriving the differential volume element $dV_{12}$ of the shifted
coordinates followed by the rotational averaging. We first obtain
\begin{equation}\label{kihacalc}
\begin{split}
dV_{12}  & = \frac{1}{3!}d^3(p_1+p_2) \\
& = \frac{1}{3!}(d^3p_1+ 3dp_2\wedge d^2p_1+(1\leftrightarrow 2))\\
& = dV_1 + \frac{1}{2} d[ p_2, dp_1, dp_1] +
(1\leftrightarrow 2)\\
& = dV_1 + H_2 dS_1 + (1\leftrightarrow 2)
\end{split}
\end{equation}
with an implicit integration in the second part and the support function $H =
pe_3$. The orientation has been chosen such that the normal surface vector of
particle $D_2$ at contact is $-e_3$. This allows to simplify the determinant,
indicated by the square brackets, via the relation $[p_2,dp_1, dp_1] =
[p,\theta^\alpha e_\alpha, \theta^\beta e_\beta]= \theta_1\wedge \theta_2 (p
e_3)$ as shown in \cite{guggenheimer}. The rotational averaging over the coset
space $\SO{3}/\SO{2}$ reduces again to the multiplication by the connection form
$\omega_1^{\;3}\wedge \omega_2^{\;3} =\kappa_1\kappa_2\theta_1\wedge \theta_2 =
K dS$: 
\begin{equation}\label{abc}
\begin{split}
\int \left<dV_{12}\right>_{\text{rot}} & = \int K_2 dS_2\wedge dV_1 \\
&\quad  + \int H_2 K_2 dS_2\wedge dS_1 + (1\leftrightarrow 2)\;.
\end{split}
\end{equation}
The product between the support function and the Gauss curvature can further
be simplified by the substitution \cite{guggenheimer}
\begin{equation}
\begin{split}
0 & = \int d[p, e_3, de_3] = \int [dp, e_3, de_3] -
[p, de_3, de_3]\\
& = 2\int (HK - M) dS\;.
\end{split}
\end{equation}
Inserting into equation (\ref{abc}), finally gives the result of Isihara and
Kihara as a special case of Minkowski's formula \cite{minkowski}
\begin{equation}
\frac{1}{4\pi}\int \left<dV_{12}\right>_{\text{rot}} = \chi_2 V_1
+ \frac{1}{4\pi} \overline{\kappa}_2 S_1 + (1\leftrightarrow 2)\;.
\end{equation}
This result can also be obtained in a coordinate-free representation by the
Lie-transport $\exp{(\mathcal{L}_{X_2})}dV_1 = dV_{12}$ of the volume form and
Stokes formula
\begin{equation}
\int_D \mathcal{L}_{X_1} \Omega_2 = \int_D d( i_{X_1} \Omega_2 ) =
\int_{\partial D} i_{X_1} \Omega_2\;.
\end{equation}
%
%
%
%
\section{}\label{sub:appendix_b}
In the following, we will give a short account of how to transform the two
particle Euler form (\ref{euler-2}) to the coordinate dependent representation
(\ref{wertheim-2}) of Wertheim, as used in \cite{wertheim-1}.

The Euclidean metric (\ref{metric}) in the orthonormal principal frame
$(\vec \nu_1, \vec \nu_2, \vec n)$ is the diagonal tensor
\begin{equation}
\begin{split}
\eta_{ij} &= e_i\otimes e_j = \mathbb{I}_{ij}\\ 
& = (\vec \nu_1\otimes \vec \nu_1 + \vec \nu_2\otimes \vec \nu_2 + \vec n\otimes
\vec n)_{ij}
\end{split}
\end{equation}
of (\ref{metric-1}). The related connection tensor (\ref{metric-1}) then follows
from the exterior derivative of the normal vector $e_3=\vec n$:
\begin{equation}
\begin{split}
de_3 &= \omega_{3\alpha}e_\alpha = \kappa_\alpha \theta_\alpha \otimes 
e_\alpha = 
\kappa_\alpha e_\alpha \otimes e_\alpha d\vec p\\
& = \bigl(\kappa_\alpha e_\alpha \otimes e_\alpha \bigr) \vec t\,ds\\
&= \bigl(\kappa_1 \vec \nu_1\otimes \vec \nu_1 + \kappa_2 \vec \nu_2\otimes
\vec \nu_2\bigr) \vec t\,ds\\
&= \mathbb{K}\;\vec t ds
\end{split}
\end{equation}
using Rodrigues formula (\ref{pre-rod}), the representation of the vielbein
$\theta_\alpha = e_\alpha d\vec p$, and by observing that the tangential vector
at each point $\vec p\in \Sigma_1\cap \Sigma_2$ lies in the direction of $\vec t
\sim \vec{n}^{(1)}\times \vec{n}^{(2)}$. The derivative $d \vec p = \vec t\,ds$
therefore is the differential line element $ds$ pointing into the direction of
$\vec t$. 

In order to separate the normal vectors from the principal frame, Wertheim
rewrites the connection form \cite{wertheim-1}:
\begin{equation}
\begin{split}
\mathbb{K} &= \frac{1}{2}\mathbb{K} + \frac{1}{2}\mathbb{K} \\
&= \frac{1}{2} \bigl(\kappa_1 \vec \nu_1\otimes \vec \nu_1 + \kappa_2
\vec \nu_2\otimes \vec \nu_2\bigr)\\ 
&\qquad + \frac{1}{2}\kappa_1 \bigl( \mathbb{I} - \vec n\otimes \vec n -
\vec \nu_2\otimes \vec \nu_2\bigr)\\
&\qquad + \frac{1}{2}\kappa_2 \bigl( \mathbb{I} - \vec n\otimes \vec n -
\vec \nu_1\otimes \vec \nu_1\bigr)\\
&= \frac{1}{2}\bigl(\kappa_1+\kappa_2)(\mathbb{I} - \vec n\otimes \vec n\bigr)\\
&\qquad + \frac{1}{2}\bigl(\kappa_1-\kappa_2)(\vec \nu_1\otimes \vec \nu_1 -
\vec \nu_2\otimes \vec \nu_2\bigr)\\
&= \bar\kappa\, \bigr(\mathbb{I} - \vec n\otimes \vec n\bigl)\; + \;\Delta
\end{split}
\end{equation}
with the mean and tangential curvatures defined in (\ref{many-curvatures}). The
connection then yields the form
\begin{equation}
\omega_{13} = e_1d e_3 =  \vec t\, \mathbb{K}\, \vec t\,ds
\end{equation}
of (\ref{euler-2}). In a second step, the normal vector $\vec n^{(2)}$ is
separated from the curvature depending parts of particle $1$:
\begin{align}\nonumber
&(\vec{n}^{(1)}\times \vec{n}^{(2)})\; \mathbb{K}_{(1)} \; (\vec{n}^{(1)}\times
\vec{n}^{(2)})\\\nonumber
&=-\vec{n}^{(2)}\times \vec{n}^{(1)}\;\bigl(\kappa_1 \vec \nu_1\otimes \vec
\nu_1 + \kappa_2 \vec \nu_2\otimes \vec \nu_2\bigr) \vec{n}^{(1)}\times
\vec{n}^{(2)}\\\nonumber
&= -\vec{n}^{(2)}\bigl( \kappa_1 \vec{n}^{(1)} \times \vec \nu_1 \otimes
\vec \nu_1\times \vec{n}^{(1)}\bigr.\\
& \qquad\quad\;\;  \bigl.+ \kappa_2 \vec{n}^{(2)} \times \vec \nu_2 \otimes
\vec \nu_2\times \vec{n}^{(1)} \bigr) \vec{n}^{(2)}\\\nonumber
&= \vec{n}^{(2)}\bigl( \kappa_1 \vec \nu_2\otimes \vec \nu_2 + 
\kappa_2 \vec \nu_1\otimes \vec \nu_1\bigr) \vec{n}^{(2)}\\\nonumber
&= \vec{n}^{(2)}\; \mathbb{K}^{\dagger}_{(1)}\; \vec{n}^{(2)}\;,\nonumber
\end{align}
using the orthonormal relation $\vec \nu_1\times \vec \nu_2 = \vec n$ and
introducing the adjoint connection tensor:
\begin{equation}
\mathbb{K}^{\dagger} = \bar\kappa\, \bigl(\mathbb{I} - \vec n\otimes
\vec n\bigr)\; - \;\Delta\;.
\end{equation}
Inserting these results into (\ref{euler-2})
\begin{align*}
&\frac{1-c_{12}}{s_{12}} \omega_{13}^{(1)}
=\frac{1-c_{12}}{s_{12}}\,\vec t\;\mathbb{K}_{(1)}\,\vec t\,ds\\
&= \frac{1-c_{12}}{s_{12}}\Bigl[
\frac{\vec{n}^{(1)}\times \vec{n}^{(2)}}{s_{12}}\mathbb{K}_{(1)} 
\frac{\vec{n}^{(1)}\times \vec{n}^{(2)}}{s_{12}}\Bigr]ds\\
&= \frac{1-c_{12}}{s_{12}^2}\vec{n}^{(2)} \mathbb{K}_{(1)}^{\dagger}
\vec{n}^{(2)} \frac{ds}{s_{12}}\\
&= \frac{1}{1+c_{12}}\vec{n}^{(2)} \mathbb{K}_{(1)}^{\dagger} \vec{n}^{(2)} 
\frac{ds}{s_{12}}\\
&= \frac{1}{1+c_{12}} \vec{n}^{(2)} \bigl[
\bar\kappa^{(1)}\bigl(\mathbb{I} - \vec{n}^{(1)}\otimes \vec{n}^{(1)}\bigr) -
\Delta^{(1)}
\bigr] \vec{n}^{(2)} \frac{ds}{s_{12}}\\
&=\frac{1}{1+c_{12}}\bigl[ \bar\kappa^{(1)} (1-c_{12}^2) -
\vec{n}^{(2)}\Delta^{(1)}\vec{n}^{(2)}\bigr] \frac{ds}{s_{12}}\\
&=\Bigl[(1-\vec{n}^{(1)}\vec{n}^{(2)})\bar\kappa^{(1)}
- \frac{\vec{n}^{(2)}\Delta^{(1)}\vec{n}^{(2)}}{1+\vec{n}^{(1)}\vec{n}^{(2)}}
\Bigr]\frac{ds}{|\vec{n}^{(1)}\times \vec{n}^{(2)}|}
\end{align*}
and using the integral representation by $\delta$-functions
\begin{equation}\label{fin-rel}
\begin{split}
&\frac{1-c_{12}}{s_{12}} \omega_{13}^{(1)}\\
&= \int_{D_1\cap D_2} \Bigl[(1-\vec{n}^{(1)}\vec{n}^{(2)})\bar\kappa^{(1)}
- \frac{\vec{n}^{(2)}\Delta^{(1)}\vec{n}^{(2)}}{1+\vec{n}^{(1)}\vec{n}^{(2)}}
\Bigr]\\
&\qquad \qquad \qquad \times  \delta(\vec{n}^{(1)}\vec r_A)
\delta(\vec{n}^{(2)}\vec r_A)
\,d^3r_A\;,
\end{split}
\end{equation}
this reproduces the first part of Wertheim's equation (\ref{wertheim-2}). The
second part follows accordingly by replacing the particle indices
$1\leftrightarrow 2$. 

The integral representation used in (\ref{fin-rel}) extends the integration
along the line element $ds$ to the entire embedding space. This and similar
relations are readily derived from the linear coordinate transformation
\begin{equation}
\eta = \vec n\vec  p\;,\;\;
\zeta = \vec m\vec p \;,\;\;
\xi = \vec e_1 x + \vec e_2 y + \vec e_3 z
\end{equation}
at the point $\vec p=(x,y,z)$ and its corresponding Jacobi determinant:
\begin{equation}
\begin{split}
d\eta\wedge d\zeta\wedge d\xi & = |\det{(\vec n, \vec m, \vec e)}|\; dx\wedge
dy\wedge dz\\[0.25em]
&= |\vec n\times \vec m|\;d^3p\;.
\end{split}
\end{equation}
Applied for the integral of an arbitrary test function $F$ and two
$\delta$-functions
\begin{align}\label{d-2}\nonumber
\int & F(\vec p)\,\delta(\vec n\vec p)\, \delta(\vec m\vec p)\, d^3p \\
&=\int F(\eta,\zeta,\xi)\; \delta(\eta)\, \delta(\zeta)\, 
\frac{d\eta\,d\zeta \,d\xi}{|\vec n\times \vec m|} \\
&=\int \widetilde F(\xi) \frac{d\xi}{|\vec n\times \vec m|}\,,\nonumber
\end{align}
it reduces to the line integral along $\xi$, as used in equation
(\ref{fin-rel}).

With one $\delta$-function included, the corresponding transformation 
\begin{equation}
\eta = \vec n\vec p\;,\;\;
\zeta = \xi = \vec e_1 x + \vec e_2 y + \vec e_3 z
\end{equation}
and $\vec e\wedge \vec e = \vec e$ yields the result:
\begin{equation}\label{d-1}
\begin{split}
\int & F(\vec p)\,\delta(\vec n\vec p)\, d^3p \\
&=\int F(\eta,\zeta,\xi)\; \delta(\eta)\,
\frac{d\eta\,d\zeta \,d\xi}{|\vec e \,\vec n|} \\
&=\int \widetilde F(\zeta, \xi) \frac{dS_n}{|\vec e\, \vec n|}
=\int \widetilde F(\zeta, \xi)\, dS
\end{split}
\end{equation}
with $\det{(\vec n, \vec e, \vec e)} = \vec e\, \vec n$ and the differential
surface element $dS_n$ in the outward pointing $\vec n$ direction. 

Analogously, the integral of three $\delta$-functions reduces to a sum of
intersection points $\{\text{pt}\}$ in the variables
\begin{equation}
\eta = \vec n\vec p\;,\;\;
\zeta = \vec m\vec p\;,\;\;
\xi = \vec l\vec p\;,
\end{equation}
\\ 
solving the algebraic equation $\eta = \zeta = \xi =0$
\begin{equation}\label{d-3}
\int F(\vec p)\,\delta(\vec n\vec p)\, \delta(\vec m\vec p)\,
\delta(\vec l\vec p)\,d^3p 
=\sum_{\{\text{pt}\}} \frac{\widetilde F(\text{pt})}{|(\vec n\times 
\vec m)\,\vec l\,|}
\end{equation}
as appears in the equation of the intersection probability of three particles
(\ref{3-euler-form}).
\bibliography{Rosenfeld_Virial}
\end{document}